\documentclass[12pt]{article}
\makeatletter
\newcommand{\singlespacing}{\let\CS=\@currsize\renewcommand{\baselinestretch}{1}\tiny\CS}
\usepackage{epsfig}
\usepackage{mathtools}

\usepackage{graphicx}
\usepackage{verbatim}   
\usepackage{color}      
\usepackage{subfigure}  
\usepackage{amsfonts}
\usepackage{amsmath}

\begin{document}
\baselineskip=24pt
\parskip = 10pt
\def \qed {\hfill \vrule height7pt width 5pt depth 0pt}
\newcommand{\ve}[1]{\mbox{\boldmath$#1$}}
\newcommand{\IR}{\mbox{$I\!\!R$}}
\newcommand{\1}{\Rightarrow}
\newcommand{\bs}{\baselineskip}
\newcommand{\esp}{\end{sloppypar}}
\newcommand{\be}{\begin{equation}}
\newcommand{\ee}{\end{equation}}
\newcommand{\beanno}{\begin{eqnarray*}}
\newcommand{\inp}[2]{\left( {#1} ,\,{#2} \right)}
\newcommand{\eeanno}{\end{eqnarray*}}
\newcommand{\bea}{\begin{eqnarray}}
\newcommand{\eea}{\end{eqnarray}}
\newcommand{\ba}{\begin{array}}
\newcommand{\ea}{\end{array}}
\newcommand{\nno}{\nonumber}
\newcommand{\dou}{\partial}
\newcommand{\bc}{\begin{center}}
\newcommand{\ec}{\end{center}}
\newcommand{\2}{\subseteq}
\newcommand{\cl}{\centerline}
\newcommand{\ds}{\displaystyle}
\newcommand{\what}{\widehat}
\def\refhg{\hangindent=20pt\hangafter=1}
\def\refmark{\par\vskip 2.50mm\noindent\refhg}

\title{\sc Analysis of Type-II Hybrid Censored Competing Risks Data}
\author{\sc Arnab Koley$^1$, Debasis Kundu$^{1,2}$ \& Ayon Ganguly$^3$}

\date{}
\maketitle
\begin{abstract}

Kundu and Gupta \cite{KG:2007}
provided the analysis of Type-I hybrid censored competing risks data, when the lifetime distribution of the competing causes
of failures follow exponential distribution.  In this paper we consider the analysis of Type-II hybrid censored competing 
risks data.  It is assumed that latent lifetime distributions of the competing causes of failures follow independent 
exponential distributions with different scale parameters.  It is observed that the maximum likelihood estimators of the unknown 
parameters do not always exist.  We propose the modified estimators of the scale parameters, which coincide with the corresponding 
maximum likelihood estimators when they 
exist, and asymptotically they are equivalent.  We obtain the exact distribution of the proposed estimators.  Using the exact distributions of the proposed estimators, associated confidence intervals are obtained.
The asymptotic and bootstrap confidence intervals of the unknown parameters are also provided. Further,  Bayesian inference of 
some unknown parametric functions under a very flexible Beta-Gamma prior is considered.  Bayes estimators and associated credible 
intervals of the unknown parameters are obtained using Monte Carlo method.   Extensive Monte Carlo simulations are 
performed to see the effectiveness of the proposed estimators and one real data set has been analyzed for the 
illustrative purposes.  It is observed that the proposed model and the method work quite well for this data set.

\end{abstract}


\noindent $^1$ Department of Mathematics and Statistics, Indian Institute of
Technology Kanpur, Pin 208016, India.  

\noindent $^2$ Corresponding author.  E-mail: kundu@iitk.ac.in, Phone no. 91-512-2597141, Fax no. 91-512-2597500.

\noindent $^3$ Department of Mathematics, Indian Institute of Technology Guwahati,  Guwahati, Pin 781039, India.


\section{\sc Introduction}

In medical or reliability analysis, it is often observed that an item can fail due to different causes.  For example, in 
a medical study it is observed that a person can die due to different diseases or in a reliability experiment it is observed 
that an automobile may fail due to different factors.  In such a situation, an investigator is often interested in the assessment 
of a specific cause, in presence of other causes.  In the statistical literature it is well known as the competing risks problem.  In a competing risk problem, 
the data consists of a failure time and an indicator denoting the causes of failure.  Among different methods, the two 
most popular approaches 
to analyze competing risks data are the following: (i) latent failure time model as suggested by Cox \cite{Cox:1959} 
or (ii) cause specific hazard functions model as suggested by Prentice et al. \cite{PAO:1978}.  
Several studies have been 
carried out over the last three decades under the above model assumptions based on both parametric and non-parametric set up.
For the parametric set up, it is assumed that lifetimes follow some specific distributions, whereas 
under the non-parametric set up no specific distributional assumptions are needed.
Interested readers are referred to the monograph of Crowder \cite{Crowder:2001} for a comprehensive review on the analysis of
different competing risks models.

Type-I and Type-II are the two most common censoring schemes which are used in practice.  
A mixture of Type-I and Type-II censoring schemes is known as hybrid censoring scheme (HCS).  
Epstein \cite{Epstein:1954} first introduced this HCS, and it is also known as
Type-I HCS.  Since the introduction of the Type-I HCS of Epstein \cite{Epstein:1954}, extensive work has been done on 
Type-I hybrid censoring schemes, see for example Fairbanks et al. \cite{FMD:1982}, Chen and Bhattacharyya \cite{CB:1988}, 
Gupta and Kundu \cite{GK:1998}, Dube et al. \cite{DPK:2011}, Kundu \cite{Kundu:2007}, and the references cited therein.
Childs et al. \cite{CCBK:2003} introduced a new hybrid censoring scheme and it can be described 
as follows.  Suppose, $n$ items are put on a test at the time point 0.  Let the ordered lifetimes of these experimental units be denoted
by $X_{1:n} < X_{2:n} < \ldots < X_{n:n}$, respectively.  Suppose $R < n$ is a pre-fixed integer and $T$ is a pre-fixed time point.
The test is terminated at a random time $T^* = \min\{X_{R:n},T\}$.  
This hybrid censoring scheme is named as the  Type-II HCS.  The main advantage of the Type-II 
HCS is that it guarantees at least $R$ failures before the end of the experiment, and if $R$ failures occur before time $T$, 
the experiment continues till time $T$, which might lead to more than $R$ failures before the experiment stops.  A detailed 
discussion on Type-I and Type-II hybrid censoring schemes can be obtained in Balakrishnan and Kundu \cite{BK:2013}.

Kundu and Gupta \cite{KG:2007} provided the analysis of Type-I hybrid censored competing risks data.  Based on the latent failure
time model assumption of Cox \cite{Cox:1959}, and the latent failure distributions to be exponential, 
the maximum likelihood estimators (MLEs) of the unknown parameters are obtained when they exist.  It is observed that the MLEs of
scale parameters may not always exist.  The exact conditional distributions of the MLEs are 
also provided.  The authors further considered the Bayesian inference of the unknown parameters based on the 
independent gamma priors, and obtained the Bayes estimates and the associated credible intervals.  For some of the recent references
in this topic, interested readers are referred to Iliopolus \cite{Ilio:2015} and Balakrishnan et al. \cite{BCI:2014}.

The aim of this paper is to provide the analysis of Type-II hybrid censored competing risks data.  In this paper also we have 
made the latent failure time model assumption of Cox \cite{Cox:1959}.  In latent failure time modeling, it is assumed that competing
causes of failures are independent random variables.  In this paper it is assumed that we have only two competing causes of failures,
and the lifetimes of the competing causes of failure follow exponential distribution with 
different scale parameters.  Therefore, if $Z$ denotes the lifetime on an item, then
$$
Z = \min\{T_1, T_2\},
$$
where $T_1$ and $T_2$ are the latent failure times of two different causes of the item.  It may be mentioned that although the 
assumption of independence of the two failure time distributions $T_1$ and $T_2$ seems to be very restrictive, it has been shown 
by Tsiatis \cite{Tsiatis:1975} that without the presence of covariates the independence between $T_1$ and $T_2$ cannot be tested
using the data only, see also Kalbfleisch and Prentice \cite{KP:1980} in this respect.  Moreover, it is observed by Kundu \cite{Kundu:2004} that in case of exponential or Weibull lifetime 
distributions, both the approaches namely the latent failure time model of Cox \cite{Cox:1959} 
or the cause specific hazard functions model of Prentice et al. \cite{PAO:1978}, provide the same likelihood function, although
their interpretations are different.

It is observed that in this case also the MLEs may not always exist.  We propose new estimators of the scale parameters which always
exist.  They coincide with the MLEs when the later exist, and asymptotically they are equivalent.  We obtain the exact distributions 
of the proposed estimators,  and it can be written as a generalized mixture of shifted gamma distributions.
 Based on the monotonicity assumption as in Chen and Bhattacharyya \cite{CB:1988}, the  confidence intervals  
based on the exact distributions are also provided.  It may be mentioned that the main purpose to propose the new estimators is to provide unconditional 
inference and also to provide confidence set of the scale parameters even though MLEs do not exist.

For comparison purposes, we have provided the asymptotic and bootstrap confidence intervals also.
We further consider the Bayesian inference of the unknown parameters.  For Bayesian analysis, we need to assume certain priors
on the unknown parameters.  In this case we have considered a very flexible Beta-Gamma prior as suggested by Pena and Gupta 
\cite{PG:1990} for the scale parameters.  The Bayes estimates can be obtained in explicit forms, and we have also provided 
joint credible set of the unknown parameters.  Extensive simulations are performed to see the effectiveness of the different 
methods, and one data set has been analyzed for illustrative purposes.

The main differences of the present paper with Kundu and Gupta \cite{KG:2007} are the following.  In Kundu and Gupta 
\cite{KG:2007}, the authors considered the analysis of Type-I hybrid censored competing risks data, whereas in this paper
we consider the analysis of Type-II hybrid censored competing risks data.  The exact distributions of the estimators  are 
quite different.  Moreover all the results available till date are based on the conditional distribution, whereas
in this paper the results are obtained without any conditioning argument.  
Using the exact distributions of the proposed estimators we have provided confidence intervals of the unknown parameters.  
It is observed that even when the MLEs do not exist, it is possible to provide
confidence set of the parameters.
Finally, in this paper we have considered the Bayesian inference of
the unknown parameters based on a very general Beta-Gamma priors, where as most of the existing results are based on independent 
gamma priors on the scale parameters which is a special case of the Beta-Gamma priors.  Based on this general prior, we have provided
the Bayes estimates and also the associated credible intervals.  We have also provided Gibbs sampling procedure to compute
the Bayes estimate of any function of unknown parameters and the associated credible interval.

Rest of the paper is organized as follows.  In Section 2, we describe the model, provide the prior assumptions, and present
the definition and notations used throughout this paper.  The proposed estimators and their exact distributions are 
derived in Section 3.  In Section 4, we have presented different confidence intervals.  Bayesian analysis has been considered
in Section 5.  Monte Carlo simulation results and the analysis of a real data set have been presented in Section 6, and finally
we conclude the paper in Section 7.

\section{\sc Notations, Model Descriptions \& Prior}

Without loss of generality, we assume that there are only two causes of failure.  We use the following notations throughout 
this paper. 

\subsection{\sc Notations}

\beanno
\hbox{PDF:} & & \hbox{probability density function}  \\
\hbox{CDF:} & & \hbox{cumulative distribution function}  \\ 
\hbox{MLE:} & & \hbox{maximum likelihood estimator} \\
\hbox{HPD:} & & \hbox{highest posterior density} \\
\hbox{MGF:} & & \hbox{moment generating function}  \\
T_{ji}: & & \hbox{latent failure time of the $i$-th individual under cause $j$, for $j$ = 1, 2}  \\
Z_i: &  & \min\{T_{1i}, T_{2i}\}  \\
Z_{i:n}: &  & \hbox{$i$-th order statistic of $Z_i$, $i = 1, \ldots, n$, and we define $Z_{n+1:n} = \infty$}  \\
T^*: &  & \max\{Z_{R:n},T\}  \\
J:&  & \hbox{the total number of failures before $T^*$}  \\
D_i: &  & \hbox{the number of failures observed due to cause $i$; $i$ = 1, 2} \\
\delta_i: &  & \hbox{indicator variable denoting the cause of failure of the $i$-th ordered individual} \\
{\cal D}: &  & \{(z_{1n},\delta_1),\ldots, (z_{Jn}, \delta_J)\}, \hbox{the observation before the experiment
stops}  \\
I_j: &  & \{z_{i:n}; \delta_i = j\}; j = 1, 2  \\
\hbox{GA}(\alpha,\lambda): & & \hbox{gamma random variable with PDF; } 
\frac{\lambda^{\alpha}}{\Gamma(\alpha)} x^{\alpha-1} e^{-\lambda x}; \ x > 0  \\
\hbox{exp}(\lambda): &  &\hbox{exponential random variable with PDF; } \lambda e^{- \lambda x}; 
x > 0, \ \ \lambda > 0,  \\
\eeanno
\beanno
\hbox{Bin}(N,p): & & \hbox{Binomial random variable with probability mass function } \\
&  & {N\choose{i}} p^i (1-p)^{N-i}, \ \ \ \hbox{    for $i = 1, 2, \ldots, N$}  \\
\hbox{Beta}(a,b): & & \hbox{Beta random variable with PDF; } \\
&  & \frac{\Gamma(a+b)}{\Gamma(a) \Gamma(b)} x^{a-1}
(1-x)^{b-1}; \  0 < x < 1  \\
f_G(x; \mu,\alpha,\lambda): & & \hbox{PDF of a shifted gamma random variable; } \\
&  &  \frac{\lambda^{\alpha}}{\Gamma(\alpha)} (x-\mu)^{\alpha-1} e^{-\lambda (x-\mu)}; \ x > \mu   \\
\overline{F}_G(x;\mu,\alpha,\lambda): & & \int_{x}^\infty f_G(z; \mu,\alpha,\lambda)dz \\
 \hbox{BG}(b_0,a_0,a_1,a_2): & & \hbox{Random variable having a Beta-Gamma distribution with}  \\
   & & \ \ \ \ \ \hbox{the PDF given by } (\ref{priors-scale}).  
\eeanno

\subsection{\sc Model Description}

Consider the following lifetime experiment in which $n$ items are put on a test.  Each unit is exposed to two risks. Here $Z_i$
denotes the lifetime of the $i$-th unit and
$$
Z_i = \min\{T_{1i}, T_{2i}\}.
$$
Based on Cox's latent failure time model assumptions, it is assumed that $T_{1i}$ and $T_{2i}$ are independently distributed for
each $i$.  Moreover, it is assumed that $T_{1i} \sim$ (follows) exp$(\lambda_1)$ and $T_{2i} \sim$ exp$(\lambda_2)$, for $i = 1, \ldots,
n$.  The test is terminated at the time point $T^*$.
It is immediate that the PDF of $Z_i$ has the following form
\begin{equation}
f(z; \lambda_1, \lambda_2)= \left (\lambda_1+\lambda_2 \right ) e^{-z \left (\lambda_1+\lambda_2 \right )};
\ \ \ \ z > 0,
\end{equation}
and 0, otherwise.  We use the following notation
$$
\lambda=\lambda_1+\lambda_2.
$$
Moreover, we have the following observation at the end of the experiment. \\
Case-I : $\{(Z_{1:n},\delta_1),\ldots,(Z_{R:n},\delta_R)\}$ ; if $T<Z_{R:n}$\\
Case-II : $\{(Z_{1:n},\delta_1),\ldots,(Z_{J:n},\delta_J)\}$ ; if $Z_{R:n}<T$ and $Z_{J:n}<T<Z_{J+1:n}$ for
$J=R,\ldots,n$.

\subsection{\sc Prior Assumption}

Borrowing the idea from Pena and Gupta \cite{PG:1990}, we assume the following joint conjugate prior on $\lambda_1$ and
$\lambda_2$.  For $a_0 > 0, a_1 > 0, a_2 > 0, b_0 > 0$, the joint prior of $(\lambda_1, \lambda_2)$ has the following PDF
for $0  <\lambda_1, \lambda_2 < \infty$.
\begin{equation}
\pi(\lambda_1,\lambda_2|b_0, a_0, a_1, a_2) =  \frac{\Gamma(a_1+a_2)}{\Gamma(a_0)} (b_0 (\lambda_1+\lambda_2))^{a_0-a_1-a_2}
\times \frac{b_0^{a_1}}{\Gamma(a_1)} \lambda_1^{a_1-1} e^{-b_0 \lambda_1}
 \times \frac{b_0^{a_2}}{\Gamma(a_2)} \lambda_2^{a_2-1} e^{-b_0 \lambda_2}.   \label{priors-scale}
\end{equation}
The joint PDF (\ref{priors-scale}) is known as the PDF of a Beta-gamma distribution.

The joint PDF (\ref{priors-scale}) can take variety of shapes.  The correlation between $\lambda_1$ and $\lambda_2$ can be both positive and
negative, depending on the values of $a_0, a_1$ and $a_2$.  If $a_0 = a_1 + a_2$, the prior distributions
of $\lambda_1$ and $\lambda_2$ become independent.
The
following results will be useful for further development.

\noindent {\sc Result 1:} If $(\lambda_1, \lambda_2) \sim$ BG($b_0,
a_0, a_1, a_2$), then for $i = 1, 2$,
 \begin{equation}
E(\lambda_i) = \frac{a_0 a_i}{b_0(a_1 + a_2)} \ \ \ \hbox{and} \ \ \
V(\lambda_i) = \frac{a_0 a_i}{b_0^2 (a_1 + a_2)} \times \left \{
\frac{(a_i+1)(a_0 + 1)}{a_1+a_2+1} - \frac{a_0 a_i}{a_1+a_2} \right
\}.     \label{result}
\end{equation}
\noindent {\sc Proof:} 
\beanno
E(\lambda_1) &=& \int_{0}^{\infty} \int_{0}^{\infty} \frac{\Gamma(a_1+a_2)}{\Gamma(a_0)} (b_0 (\lambda_1+\lambda_2))^{a_0-a_1-a_2}
\times \frac{b_0^{a_1}}{\Gamma(a_1)} \lambda_1^{a_1} e^{-b_0 \lambda_1}
 \times \frac{b_0^{a_2}}{\Gamma(a_2)} \lambda_2^{a_2-1} e^{-b_0 \lambda_2} d\lambda_1 d\lambda_2\\
 &=& \frac{\Gamma(a_0+1)}{\Gamma(a_1+1+a_2)} \frac{\Gamma(a_1+a_2)}{\Gamma(a_0)} \frac{\Gamma(a_1+1)}{\Gamma(a_1)} \frac{1}{b_0}\\
 &  & \int_{0}^{\infty} \int_{0}^{\infty} \frac{\Gamma(a_1+1+a_2)}{\Gamma(a_0+1)} (b_0 (\lambda_1+\lambda_2))^{a_0+1-a_1-1-a_2}
\times \frac{b_0^{a_1+1}}{\Gamma(a_1+1)} \lambda_1^{a_1} e^{-b_0 \lambda_1}
 \times \frac{b_0^{a_2}}{\Gamma(a_2)} \lambda_2^{a_2-1}\\
& &\quad \quad \quad e^{-b_0 \lambda_2} d\lambda_1 d\lambda_2\\
&=& \frac{a_0a_1}{b_0(a_1+a_2)}.
\eeanno
The integration in the second step of the above turns out to be 1 because of equation (\ref{priors-scale}). Similarly we get 
$\ds E(\lambda_2)=\frac{a_0a_2}{b_0(a_1+a_2)}$.
Next we derive $V(\lambda_1)$. Note that $V(\lambda_1)=E(\lambda_1^2)-(E(\lambda_1))^2$. We derive $E(\lambda_1^2)$ below.
\beanno
E(\lambda_1^2)&=& \int_{0}^{\infty} \int_{0}^{\infty} \frac{\Gamma(a_1+a_2)}{\Gamma(a_0)} (b_0 (\lambda_1+\lambda_2))^{a_0-a_1-a_2}
\times \frac{b_0^{a_1}}{\Gamma(a_1)} \lambda_1^{a_1+1} e^{-b_0 \lambda_1}
 \times \frac{b_0^{a_2}}{\Gamma(a_2)} \lambda_2^{a_2-1} e^{-b_0 \lambda_2} d\lambda_1 d\lambda_2\\
 &=& \frac{\Gamma(a_1+2)}{\Gamma(a_1)} \frac{\Gamma(a_1+a_2)}{\Gamma(a_0)} \frac{\Gamma(a_0+2)}{\Gamma(a_1+a_2+2)} \frac{1}{b^2_0}\\
 &=& \int_{0}^{\infty} \int_{0}^{\infty} \frac{\Gamma(a_1+2+a_2)}{\Gamma(a_0+2)} (b_0 (\lambda_1+\lambda_2))^{a_0+2-a_1-2-a_2}
\times \frac{b_0^{a_1+2}}{\Gamma(a_1+2)} \lambda_1^{a_1} e^{-b_0 \lambda_1}
 \times \frac{b_0^{a_2}}{\Gamma(a_2)} \lambda_2^{a_2-1}\\
 & & \quad \quad \quad e^{-b_0 \lambda_2} d\lambda_1 d\lambda_2\\
 &=& \frac{a_0a_1(a_0+1)(a_1+1)}{b^2_0(a_1+a_2)(a_1+a_2+1)}.
\eeanno
Thus 
\beanno
V(\lambda_1) & = & \frac{a_0 a_1}{b_0^2 (a_1 + a_2)} \times \left \{\frac{(a_1+1)(a_0 + 1)}{a_1+a_2+1} - \frac{a_0 a_1}{a_1+a_2} \right
\},    \\
V(\lambda_2) & = & \frac{a_0 a_2}{b_0^2 (a_1 + a_2)} \times \left \{
\frac{(a_1+1)(a_0 + 1)}{a_1+a_2+1} - \frac{a_0 a_2}{a_1+a_2} \right
\}.
\eeanno
Next we provide steps to generate samples from $BG(b_0,a_0,a_1,a_2)$ distribution.  We need the following Lemma.  The proof is 
quite straight forward, hence the details are avoided.

\noindent {\sc Lemma 1:} If $\ds (\lambda_1,\lambda_2)\sim BG(b_0,a_0,a_1,a_2)$ then
$$
U = \lambda_1+\lambda_2 \sim \hbox{GA} (a_0,b_0) \ \ \  \hbox{and} \ \ \  V = \frac{\lambda_1}{\lambda_1+\lambda_2} \sim \hbox{Beta} (a_1,a_2).
$$
Moreover, $U$ and $V$ are independent.

Using the same algorithm as suggested in Kundu and Pradhan \cite{KP:2011} following steps are required to generate samples from a Beta-Gamma distribution.
\begin{itemize}
\item{\sc Step-1} Generate  $\ds U$ from $\text{GA} (a_0,b_0)$.
\item{\sc Step-2} Generate $\ds V$ from $\ds \text{Beta}(a_1,a_2)$.
\item{\sc Step-3} Obtain $\ds \lambda_1=UV $ and $\ds \lambda_2=U(1-V).$
\end{itemize}

\section{\sc Estimators of $\lambda_1$ and $\lambda_2$ and their Distributions}

\subsection{\sc Estimators of $\lambda_1$ and $\lambda_2$}

\noindent Likelihood contribution of the data point $(z,\delta = j)$, for $j = 1, 2$, is given by 
\[L(\lambda_1,\lambda_2|(z,\delta=j))=\lambda_j e^{-\lambda_1 z} e^{-\lambda_2 z}=\lambda_j e^{-(\lambda_1+\lambda_2) z}. \]
Thus, the likelihood function of the observation is given by
\[ L(\lambda_1,\lambda_2|Data) = \left\{
  \begin{array}{l l}
    \frac{n!}{D_1! D_2!(n-R)!}\lambda_1^{D_1}\lambda_2^{D_2}e^{-W(\lambda_1+\lambda_2)} \quad \text{if $T<Z_{R:n}$}\\
    \frac{n!}{D_1! D_2!(n-J)!}\lambda_1^{D_1}\lambda_2^{D_2}e^{-W(\lambda_1+\lambda_2)} \quad \text{if $Z_{J:n}<T<Z_{J+1:n}$},
  \end{array} \right.\]
here $R \le J \le n$.  Hence, for
\[
W = \left \{ \ba{l r r} \sum_{i=1}^R Z_i+Z_R(n-R) \quad \hbox{for} \quad \hbox{Case-I}  \\
\sum_{i=1}^J Z_i+T(n-J) \quad \hbox{\ \ for} \quad \hbox{Case-II,}  \ea  \right .
\]
the log likelihood function without the additive constant is
\begin{equation}
l(\lambda_1,\lambda_2|Data)=D_1 \ln \lambda_1 + D_2 \ln \lambda_2 - W \big(\lambda_1+\lambda_2\big).
\end{equation}
Clearly, the MLEs of $\lambda_1$ and $\lambda_2$ are given by 
$$\what{\lambda}_{1MLE} = \frac{D_1}{W} \ \ \ \hbox{if} \ \ D_1>0 \ \ \ \ \ \hbox{and} \ \ \ \what{\lambda}_{2MLE} = \frac{D_2}{W} 
\ \ \hbox{if} \ \ \ D_2 > 0.
$$ 
Note that when $D_1=0$, the MLE of $\lambda_1$ does not exist, and similarly, 
when $D_2$ = 0, the MLE of $\lambda_2$ does not exist.  We define the estimators of $\lambda_1$ and $\lambda_2$ which will
be useful for constructing their confidence intervals even when the MLEs do not exist.  The proposed estimators are as follows:
$$
\what{\lambda}_1 = \left \{ \begin{matrix}\what{\lambda}_{1MLE} & \hbox{if} & D_1 > 0  \cr 
0 & \hbox{if} & D_1=0 \end{matrix}  \right . \ \ \ \ \hbox{and} \ \ \ \
\what{\lambda}_2 = \left \{ \begin{matrix}\what{\lambda}_{2MLE} & \hbox{if} & D_2 > 0  \cr 
0 & \hbox{if} & D_2=0. \end{matrix}  \right .
$$ 
Therefore, although the MLEs of $\lambda_1$ and $\lambda_2$ may not always exist, $\what{\lambda}_1$ and $\what{\lambda}_2$ 
always exist, see for example the definition of an estimator in Definition 7.1.1 of Casella and Berger \cite{CB:1990}.  
Now we obtain the exact distributions of $\what{\lambda}_1$ and $\what{\lambda}_2$, and based on the exact 
distributions of $\what{\lambda}_1$ and $\what{\lambda}_2$, exact confidence intervals can be constructed.  We will also show
that even when $D_1$ = 0 or $D_2$ = 0, it is possible to obtain the exact confidence set of $(\lambda_1, \lambda_2)$.

\subsection{\sc Distributions of $\what{\lambda}_1$ and $\what{\lambda}_2$}

In this section we provide the exact distributions of $\what{\lambda}_1$ and $\what{\lambda}_2$.
For $x \geq 0$, the distribution function of $\what{\lambda}_1$ is given
\bea
P(\what{\lambda}_1 \leq x) & = & P(\what{\lambda}_1 \leq x| D_1=0) P(D_1=0)+ P(0 < \what{\lambda}_1 \leq x| D_1>0)P(D_1>0) \nonumber \\
& = & P(D_1=0)+ P(0 < \what{\lambda}_1 \leq x| D_1>0)P(D_1>0)\nonumber \\
& = & \sum \limits_{i=0}^{n} c_i + \sum \limits_{i=1}^R \sum \limits_{s=0}^{R-1} c_{i,s}(x) + \sum \limits_{j=R}^n \sum \limits_{i=1}^{j} \sum \limits_{s=0}^{j} d_{j,i,s}(x), \label{dist-1}
\eea
where\\
(a)
\noindent
\begin{align*}
c_i  = \begin{dcases}
 \binom{n}{i} (1-e^{-T\lambda})^i e^{-(n-i)T\lambda} \Big(\frac{\lambda_2}{\lambda_1+\lambda_2}\Big)^R, &  i=0,1,\ldots R-1\\
 \binom{n}{i} (1-e^{-T\lambda})^i e^{-(n-i)T\lambda} \Big(\frac{\lambda_2}{\lambda_1+\lambda_2}\Big)^i, & i=R,R+1,\ldots n,\\ 
 \end{dcases}  
\end{align*}
(b) for $i=1,\ldots, R \quad \text{and} \quad s=0,\ldots, R-1,$
\begin{align*}
c_{i,s}(x) & = n \binom{n-1}{R-1} \binom{R-1}{s} \binom{R}{i} \Big(\frac{\lambda_1}{\lambda_1+\lambda_2}\Big)^i \Big(\frac{\lambda_2}{\lambda_1+\lambda_2}\Big)^{R-i}  \frac{(-1)^s}{(n-R+s+1)}e^{-T\lambda (n-R+1+s)}\\
& \quad \overline{F}_G \bigg(\frac{1}{x};\frac{T}{i}\big(n-R+s+1\big),R,i \lambda \bigg),
\end{align*}
(c.) for $s=0,\ldots,j,\quad  i=1, \ldots, j\quad  \text{and}\quad  j=R,\ldots, n,$
\begin{align}
d_{j,i,s}(x) & = \binom{n}{j} \binom{j}{i} \binom{j}{s} \Big(\frac{\lambda_1}{\lambda_1+\lambda_2}\Big)^i \Big(\frac{\lambda_2}{\lambda_1+\lambda_2}\Big)^{j-i}  (-1)^s e^{-T\lambda (n-j+s)}\overline{F}_G \bigg(\frac{1}{x};\frac{T}{i}\big(n-j+s \big),j,i \lambda \bigg). \label{distribution}
\end{align}
\noindent {\sc Proof:} See in the Appendix.   \qed

It is clear that the distribution of $\what{\lambda}_1$ is  mixture of a degenerate and an absolute continuous distributions.  The 
PDF of the absolute continuous part of the distribution of $\what{\lambda}_1$ can be written as
$$
f_{\what{\lambda}_1|D_1 > 0}(x) = \frac{1}{P(D_1 > 0)} \left [ \sum \limits_{i=1}^R \sum \limits_{s=0}^{R-1} \frac{d}{dx}c_{i,s}(x) + 
\sum \limits_{j=R}^n \sum \limits_{i=1}^{j} \sum \limits_{s=0}^{j} \frac{d}{dx} d_{j,i,s}(x)  \right ],
$$
where,\\
(i) for $i=1,\ldots, R \quad \text{and} \quad s=0,\ldots, R-1,$
\begin{align*}
\frac{d}{dx} c_{i,s} (x) & = n \binom{n-1}{R-1} \binom{R-1}{s} \binom{R}{i} \Big(\frac{\lambda_1}{\lambda_1+\lambda_2}\Big)^i \Big(\frac{\lambda_2}{\lambda_1+\lambda_2}\Big)^{R-i}  \frac{(-1)^s}{(n-R+s+1)}\\
& \quad e^{-T\lambda (n-R+1+s)} \frac{1}{x^2} f_G \bigg(\frac{1}{x};\frac{T}{i}\big(n-R+s+1\big),R,i \lambda \bigg),
\end{align*}
(ii) for $s=0,\ldots,j,\quad  i=1, \ldots, j\quad  \text{and}\quad  j=R,\ldots, n,$
\begin{align}
\frac{d}{dx} d_{j,i,s} (x)  & = \binom{n}{j} \binom{j}{i} \binom{j}{s} \Big(\frac{\lambda_1}{\lambda_1+\lambda_2}\Big)^i \Big(\frac{\lambda_2}{\lambda_1+\lambda_2}\Big)^{j-i}  (-1)^s e^{-T\lambda (n-j+s)}\nonumber \\
& \quad \frac{1}{x^2} f_G \bigg(\frac{1}{x};\frac{T}{i}\big(n-j+s \big),j,i \lambda \bigg). \label{density}
\end{align}
Similarly, the distribution function of $\what{\lambda}_2$ and the PDF of the absolute continuous part of the distribution of 
$\what{\lambda}_2$ can obtained from equations (\ref{distribution}) and (\ref{density}), respectively, by interchanging 
$\lambda_1$ and $\lambda_2$.

\noindent {\sc Comment:} Note that if we denote $\theta_1$ = $1/\lambda_1$ and $\theta_2 = 1/\lambda_2$, then the MLEs of $\theta_1$ 
and $\theta_2$ exist, if $D_1 > 0$ and $D_2 > 0$, respectively.  The MLE of $\theta_1$, say $\widehat{\theta}_1$, given $D_1 > 0$ is 
$1/\widehat{\lambda}_1$ given $D_1 > 0$.  Similarly, the MLE of $\theta_2$ given $D_2 > 0$ is $\widehat{\theta}_2 = 1/\widehat{\lambda}_2$ 
given $D_2 > 0$.  Therefore, the distributions of $\widehat{\theta}_1$ given $D_1 > 0$ and $\widehat{\theta}_2$ given $D_2 > 0$ can be
easily obtained from (\ref{dist-1}).

\section{\sc Confidence interval}

In this section we present different methods of constructing 100(1-$\alpha$)\% confidence interval of $\lambda_1$.  Similar
methods can be applied to construct confidence interval of $\lambda_2$ also, and they are not presented here.
\subsection{\sc Confidence Intervals Based on Exact Distributions}

First we consider the case when $D_1 > 0$ and $D_2 > 0$.  The case when $D_1 = 0$ or $D_2 = 0$, will be discussed later.
The method of  construction of the confidence interval of $\lambda_1$ based on the 
exact conditional distribution of $\widehat{\lambda}_1$ is based on a 
similar assumption as those of Chen and Bhattacharyya \cite{CB:1988} or Kundu and Basu \cite{KB:2000}.
First let us assume that $\lambda_2$ is known.  Suppose 
$\ds P_{\lambda_1}(\widehat{\lambda}_1 \ge b)$ is a strictly increasing function of $\lambda_1$ for all $b > 0$, 
 and let $b(\cdot)$ be a 
function such that $\ds  P_{\lambda_1}(\widehat{\lambda}_1 \ge b(\lambda_1)) = \frac{\alpha}{2}$.  Therefore, for $\lambda_1 < \lambda_1'$
\be
\frac{\alpha}{2} = P_{\lambda_1'} \left [ \widehat{\lambda}_1 \ge b(\lambda_1') \right ] = 
P_{\lambda_1} \left [ \widehat{\lambda}_1 \ge b(\lambda_1) \right ] < P_{\lambda_1'} \left [ \widehat{\lambda}_1 \ge b(\lambda_1) \right ],
    \label{eq1-ci}
\ee
which implies $b(\cdot)$ is an strictly increasing function as $b(\lambda_1) < b(\lambda_1')$.  Hence $b^{-1}(\lambda)$ exists and it is also an 
increasing function.  Now from (\ref{eq1-ci}), we have
\be
1 - \frac{\alpha}{2} = P_{\lambda_1} \left [ \widehat{\lambda}_1 \le b(\lambda_1) \right ]  = P_{\lambda_1} \left [ b^{-1}(\what{\lambda}_1) \le \lambda_1 \right ].  \label{eq2-ci}
\ee
Clearly, (\ref{eq2-ci}) indicates that $\ds b^{-1}(\what{\lambda}_1)$ is the symmetric lower bound of the 100($1-\alpha$)\% 
confidence intervals of $\lambda_1$.  Therefore, if $\ds \what{\lambda}_{1,obs}$ denotes the observed value of $\what{\lambda}_1$, then 
we need to find $\ds \lambda_{1L} = b^{-1}(\what{\lambda}_{1,obs})$, such that
\be
\frac{\alpha}{2} = P_{\lambda_{1L}} (\what{\lambda}_1 \ge \what{\lambda}_{1,obs}).     \label{eq3-ci}
\ee
Note that, (\ref{eq3-ci}) is equivalent in finding 
\be
1 - \frac{\alpha}{2} = P_{\lambda_{1L}} (\what{\lambda}_1 \le \what{\lambda}_{1,obs}).     \label{eq4-ci}
\ee
Similarly, we can obtain $\lambda_{1U}$, the symmetric upper bound of the 100($1-\alpha$)\% confidence intervals of 
$\lambda_1$ by solving the following equation
\be
\frac{\alpha}{2} = P_{\lambda_{1U}} (\what{\lambda}_1 \le \what{\lambda}_{1,obs}).     \label{eq5-ci}
\ee
Since, it is not possible to obtain a closed form expression of $b(\lambda)$, we need to use some iterative method to solve 
(\ref{eq4-ci}) and (\ref{eq5-ci}) to compute $\lambda_{1L}$ and $\lambda_{1U}$, respectively.  In practice since $\lambda_2$ is 
also unknown we replace it by its MLE.

\begin{figure}[ht!]
\centering
\includegraphics[height =6.5cm,width=90mm]{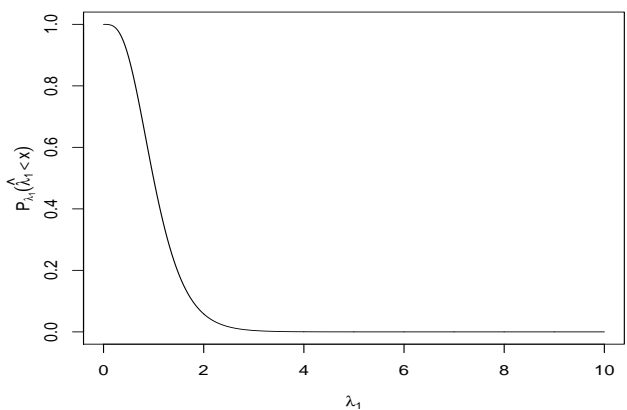}
\caption{Plot of $P_{\lambda_1}(\what{\lambda}_1 \le x)$, for $n=10, R=8, T=1.2, \lambda_2=1.3, x=1.0$}
\label{fig-1}
\end{figure}

\begin{figure}[ht!]
\centering
\includegraphics[height = 6.5cm, width=90mm]{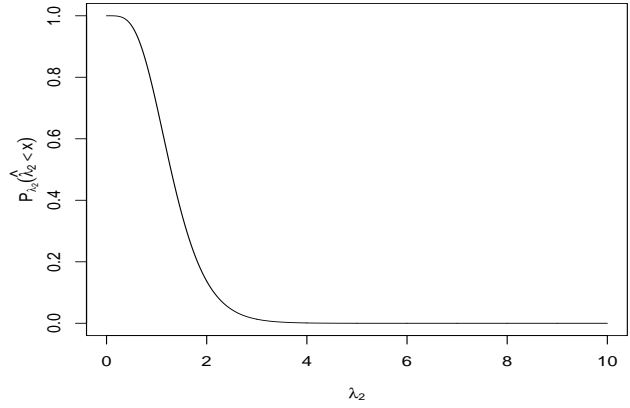}
\caption{Plot of $P_{\lambda_2}(\what{\lambda}_2 \le x)$, for $n=10, R=8, T=1.2, \lambda_1=1.0, x=1.3$}
\label{fig-2}
\end{figure}

The construction of the confidence interval of $\lambda_1$ is based on the assumption that 
$\ds P_{\lambda_1}(\widehat{\lambda}_1 \ge b)$ is a strictly increasing function of $\lambda_1$ for all $b > 0$.  Unfortunately, due to complicated 
nature of $\ds P_{\lambda_1}(\widehat{\lambda}_1 \ge b)$, we could not establish this property.  It may be mentioned that 
many authors including Chen and Bhattacharyya \cite{CB:1988}, Gupta and Kundu \cite{GK:1998}, Childs  et al. 
\cite{CCBK:2003} used this property to find confidence interval of the scale parameter for exponential distribution.  
Although theoretically it is difficult to check the assumption, a numerical study supports the monotonicity assumption.  
We present the graphs $P_{\lambda_i}(\hat{\lambda}_i\leq x)$ for $i=1,2$ in Figure \ref{fig-1} and Figure \ref{fig-2}, and  they 
support our claim.  Moreover, heuristically it may be argued that since $\lambda_1$ is a scale parameter, the distribution function
of $\widehat{\lambda}_1$ is stochastically increasing as a function of $\lambda_1$.  That justifies the assumption.
Based on the assumption that 
$\ds P_{\lambda_1}(\widehat{\lambda}_1 \ge b)$ is a strictly increasing function of $\lambda_1$ for all $b > 0$ we have the following 
result.

\noindent {\sc Lemma 5:} For $D_1 > 0$ and $D_2 > 0$, the solutions of (\ref{eq4-ci}) and (\ref{eq5-ci}) always exist.

\noindent {\sc Proof:} See in the Appendix.   \qed

Now let us consider the case when either $D_1$ = 0 or $D_2$ = 0.  Note that when $D_1$ = 0, $D_2 \ge R$, and vice versa.  Now when
$D_1$ = 0, a 100$(1-\alpha)\%$ confidence set of $(\lambda_1, \lambda_2)$ can be obtained as follows:
$$
\textbf{A} = \{(\lambda_1, \lambda_2): P_{\lambda_1, \lambda_2}(D_1=0) > 1-\alpha\}.
$$
Similarly, when $D_2$ = 0, a 100$(1-\alpha)\%$ confidence set of $(\lambda_1, \lambda_2)$ can be obtained as follows:
$$
\textbf{B} = \{(\lambda_1, \lambda_2): P_{\lambda_1, \lambda_2}(D_2=0) > 1-\alpha\}.
$$

\subsection{\sc Asymptotic and Bootstrap Confidence Intervals}

Since the construction of the confidence intervals of $\lambda_1$ and $\lambda_2$ based on the exact distributions of the estimators are
quite computationally involved, we propose to use two alternative confidence intervals which can be obtained more conveniently.
Based on the asymptotic normality of the MLEs,
100($1-\alpha$)\% asymptotic confidence interval of $\lambda_1$ and $\lambda_2$ can be obtained as
\be
\left (\what{\lambda}_1-z_{\frac{\alpha}{2}}\frac{D_1^{1/2}}{W},  \hat{\lambda}_1+z_{\frac{\alpha}{2}}\frac{D_1^{1/2}}{W} \right )
\ \ \ \hbox{and} \ \ \ \left (\what{\lambda}_2-z_{\frac{\alpha}{2}}\frac{D_2^{1/2}}{W},
\what{\lambda}_2+z_{\frac{\alpha}{2}}\frac{D_2^{1/2}}{W} \right ),
\ee
respectively. Note that asymptotic interval of $\lambda_1$ does not exist when $D_1=0$. Similar case holds for $\lambda_2$ also. 

We propose to use bootstrap method for constructing confidence intervals of the unknown parameters.  Steps of construction of
bootstrap confidence intervals are as follows.
\begin{itemize}
\item {\sc Step 1:} Define,\\
\begin{eqnarray*}
\what{\lambda}_{1M}=
    \begin{cases}
        \what{\lambda}_{1MLE} \quad \text{if $D_1>0$}\\
        \lambda_{11} \quad \quad ~ \text{if $D_1=0$ and $\lambda_{11} \in$ \textbf{A}}
    \end{cases}
    \what{\lambda}_{2M}=
    \begin{cases}
        \what{\lambda}_{2MLE} \quad \text{if $D_2>0$}\\
        \lambda_{22} \quad \quad ~ \text{if $D_2=0$ and $\lambda_{22} \in$ \textbf{B}.}
    \end{cases}
\end{eqnarray*}
Here $\lambda_{11}$ and $\lambda_{22}$ are obtained from the following two non-linear equations:
$$
P_{\lambda_{11},\widehat{\lambda}_2}(D_1 = 0) = 0.5 \ \ \ \hbox{and} \ \ \ 
P_{\widehat{\lambda}_1, \lambda_{22}}(D_2 = 0) = 0.5.
$$
\item {\sc Step 2:} \\
\textbf{Case-I}: $T<z_{R:n}$
\begin{itemize}
\item [{(a)}] Generate a sample of size $R$ from the distribution
$$f(x)=\frac{\what{\lambda}_M e^{-\what{\lambda}_M x}}{1-e^{-\what{\lambda}_M Z_{R:n}}}\ \ \ \ \ \ 0<x<z_{R:n},$$
where $\ds \what{\lambda}_{M}=\what{\lambda}_{1M}+\what{\lambda}_{2M}$.
If the largest value of the sample is greater than $T$, perform Step (b), otherwise, repeat Step (a).
\item [{(b)}] Assign Cause-I or Cause-II to each failure with probability $\ds \frac{\what{\lambda}_{1M}}{\what{\lambda}_{1M}+\what{\lambda}_{2M}}$ and \linebreak $\ds \frac{\what{\lambda}_{2M}}{\what{\lambda}_{1M}+\what{\lambda}_{2M}}$, respectively.
\end{itemize}
\textbf{Case-II}: $T > z_{R:n}$
\begin{itemize}
\item [{(a)}] Generate a sample of size $\ds J=R,R+1,\ldots,n$ from the distribution
$$f(x)=\frac{\what{\lambda}_M e^{-\what{\lambda}_M x}}{1-e^{-\what{\lambda}_M T}}\ \ \ \ \ \ 0<x<T.$$
\item [{(b)}] Assign Cause-I or Cause-II to each failure with probability $\ds \frac{\what{\lambda}_{1M}}{\what{\lambda}_{1M}+\what{\lambda}_{2M}}$ and \linebreak $\ds \frac{\what{\lambda}_{2M}}{\what{\lambda}_{1M}+\what{\lambda}_{2M}}$, respectively.
\end{itemize}
\item {\sc Step 3:} Compute $\what{\lambda}^*_{1M}$ and $\what{\lambda}^*_{2M}$ from the bootstrap sample. Repeat the process $N$ times.
\item {\sc Step 4:} Let $\what{CDF}(x)$ be the empirical distribution function of $\what{\lambda}_{1M}$.
Let us define $\what{\lambda}_{1M,boot}(x)=\what{CDF}^{-1}(x)$.  Then approximate $100(1-\alpha)\%$ confidence interval of
$\lambda_1$ is given by $\Big(\what{\lambda}_{1M,boot}(\frac{\alpha}{2}),\what{\lambda}_{1M,boot}(1-\frac{\alpha}{2})\Big)$. Similarly, we
can obtain the bootstrap confidence interval of $\lambda_2$ also.
\end{itemize}

\section{\sc Bayesian analysis}

It is assumed that $(\lambda_1, \lambda_2)$ has a joint Beta-Gamma prior as given in (\ref{priors-scale}).  Now based on the 
above joint prior, we provide the Bayes estimates and the associated credible set of the unknown parameters.  The joint 
posterior distribution of $\lambda_1$ and $\lambda_2$ can be easily observed as 
\be
\pi(\lambda_1,\lambda_2|data) \propto e^{-(W+b_0)(\lambda_1+\lambda_2)}{\lambda_1}^{a_1+D_1-1}{\lambda_2}^{a_2+D_2-1} 
{(\lambda_1+\lambda_2)}^{a_0-a_1-a_2},~~\lambda_1>0, \lambda_2>0.     \label{post}
\ee
Hence,
\be
\pi(\lambda_1,\lambda_2|data) \sim \hbox{BG}(b_0+W, a_0+J, a_1+D_1, a_2+D_2).
\ee
Therefore, under the squared error loss function the Bayes estimates of $\lambda_1$ and $\lambda_2$ are 
\begin{equation*}
\begin{split}
&\hat{\lambda}_{1B}=E(\lambda_1|data)=\frac{(a_0+J)(a_1+D_1)}{(b_0+W)(a_1+a_2+J)},\\
&\hat{\lambda}_{2B}=E(\lambda_2|data)=\frac{(a_0+J)(a_2+D_2)}{(b_0+W)(a_1+a_2+J)}, \\
\end{split}
\end{equation*}
respectively.  
Similarly, we can obtain the corresponding posterior variances as follows;
\begin{equation*}
\begin{split}
V(\lambda_1|data)=A_1B_1, ~~~~  V(\lambda_2|data)=A_2B_2 .~~~~~~~~~~
\end{split}
\end{equation*}
Here, for $k=1,2$,
\begin{equation*}
\begin{split}
&A_k=\frac{(a_0+J)(a_k+D_k)}{(b_0+W)^2(a_1+a_2+J)} \ \ \ \ \ \ \hbox{and} \\
&B_k= \frac{(a_0+J+1)(a_k+D_k+1)}{a_1+a_2+J+1}
-\frac{(a_0+J)(a_k+D_k)}{a_1+a_2+J}.
\end{split}
\end{equation*}
First we will provide a joint 100($1-\alpha$)\% credible set of $(\lambda_1, \lambda_2)$.  Let us recall that  
$C_\alpha$ is said to be a 100($1-\alpha$)\% credible set of ($\lambda_1,\lambda_2$) if,
\[P((\lambda_1,\lambda_2)\in C_\alpha)=1-\alpha,\ \ \text{where}\ \ (\lambda_1,\lambda_2)\sim \pi(\lambda_1,\lambda_2|data) .\]
Using Lemma 1, $C_\alpha$ can be constructed as follows.  First, let us choose $\alpha_1$ and $\alpha_2$, such 
that $(1-\alpha) = (1-\alpha_1)(1-\alpha_2)$.  Now, choose $C_{\alpha}$, as follows:
\be
C_\alpha= \left \{(\lambda_1,\lambda_2):\lambda_1>0,\lambda_2>0,A\leq \lambda_1+\lambda_2\leq B,C\leq \frac{\lambda_1}{\lambda_1+\lambda_2}\leq D \right \}.      \label{cre-set}
\ee
Here $A,B,C,D$ are such that 
$$
P(A\leq U \leq B)=1-\alpha_1 \ \ \ \ \hbox{and} \ \ \  
P(C\leq V \leq D)=1-\alpha_2.
$$
Note that, $C_{\alpha}$ is a trapezoid enclosed by the following four straight lines:
\be
(i) \ \lambda_1+\lambda_2 = A, \ \ (ii) \ \lambda_1+\lambda_2 = B, \ \ (iii) \ \lambda_1(1-D) = \lambda_2 D, \ \ (iv) \ \lambda_1(1-C) 
= \lambda_2 C.   \label{trapizoid}
\ee
The area of the trapezoid (\ref{trapizoid}) is $(B^2 - A^2)(D-C)/2$.

Kundu and Pradhan \cite{KP:2011} provided a very efficient algorithm to generate samples from a Beta-Gamma distribution.  Now we
will provide two algorithms for (i) computing the Bayes estimates of any function of $\lambda_1$ and $\lambda_2$, say 
$g(\lambda_1, \lambda_2)$, and also to construct HPD credible interval of $g(\lambda_1, \lambda_2)$, (ii) constructing the credible 
set of $(\lambda_1, \lambda_2)$ as described above.

\noindent {\sc Algorithm 1:} Construct Bayes estimate of $g(\lambda_1, \lambda_2)$, and the associated 100($1-\alpha$)\% 
HPD credible interval.

\begin{itemize}
\item Step-1: Generate $\lambda_1$ and  $\lambda_2$ from $ \pi(\lambda_1,\lambda_2|data)$ using the method suggested by 
Kundu and Pradhan \cite{KP:2011}.

\item Step-2: Repeat Step-1 $M$ times to generate $\ds (\lambda_{11},\lambda_{21}),\ldots,(\lambda_{1M},\lambda_{2M})$.

\item Step-3: The Bayes estimate of $\ds g(\lambda_1,\lambda_2)$ and the corresponding posterior variance can be obtained as
$$
\what{g}(\lambda_1,\lambda_2) = \frac{1}{M} \sum \limits_{i=1}^M g(\lambda_{1i},\lambda_{2i}) \ \ \  \hbox{and} \ \ \
\what{V}(g(\lambda_1,\lambda_2))=\frac{1}{M}\sum\limits_{i=1}^M (g(\lambda_{1i},\lambda_{2i})-\hat{g}(\lambda_1,\lambda_2))^2.
$$

\item Step-4: To construct credible interval of $\ds g(\lambda_1,\lambda_2)$, first order $\ds g_i$ as 
$\ds g_{(1)}<\ldots <g_{(M)}$, where $\ds g_i=g(\lambda_{1i},\lambda_{2i})$.  Then a $100(1-\alpha)\%$ credible interval 
of $g(\lambda_1,\lambda_2)$ becomes
$$
(g_{(j)},g_{(j+M(1-\alpha))}) 
$$
for $j=1,\ldots,[M\alpha]$.

\item Step-5: $100(1-\alpha)\%$ HPD credible interval becomes $\ds (g_{(j^*)},g_{(j^*+M(1-\alpha))})$ where $j^*$ is such that 
$$
g_{(j^*+M(1-\alpha))}-g_{(j^*)}\leq g_{(j+M(1-\alpha))}-g_{(j)}, \ \ \ \forall j=1,\ldots, [M\alpha].
$$

\end{itemize}

\noindent {\sc Algorithm 2:} Construction of the credible set $C_{\alpha}$ as given in (\ref{cre-set}). 

\begin{itemize}
\item Step-1: Generate $\lambda_1$ and  $\lambda_2$ from $ \pi(\lambda_1,\lambda_2|data)$ using the method suggested by 
Kundu and Pradhan \cite{KP:2011}.

\item Step-2: Repeat Step-1 $M$ times to generate $\ds (\lambda_{11},\lambda_{21}),\ldots,(\lambda_{1M},\lambda_{2M})$.

\item Step-3: Compute for $i = 1, \ldots, M$,
$$
u_i = \lambda_{1i}+\lambda_{2i} \ \ \  \hbox{and} \ \ \ v_i = \frac{\lambda_{1i}}{\lambda_{1i}+\lambda_{2i}}. 
$$

\item Step-4: Order $u_1, \ldots, u_M$, as $u_{(1)} < \ldots < u_{(M)}$, similarly, order $v_1, \ldots, v_M$, as 
$v_{(1)} < \ldots < v_{(M)}$.

\item Step-5: Now compute $A = u_{(j^*)}$ and $B = u_{(j^*+M(1-\alpha_1))}$, where
$$
u_{(j^*+M(1-\alpha))}^2-u_{(j^*)}^2 \leq u_{(j+M(1-\alpha))}^2 - u_{(j)}^2, \ \ \ \forall j=1,\ldots, [M\alpha].
$$

\item Step-6: Similarly, $C = v_{(j^*)}$ and $D = v_{(j^*+M(1-\alpha_1))}$, where
$$
v_{(j^*+M(1-\alpha))}-v_{(j^*)} \leq v_{(j+M(1-\alpha))} - v_{(j)}, \ \ \ \forall j=1,\ldots, [M\alpha].
$$

\end{itemize}

\section{\sc Simulation Results}

In this section we present some simulation results to see how different methods behave for small sample sizes and also 
for different parameter values.   We compare the performances of the proposed estimators and  
the Bayes estimators both under informative and non-informative priors.  We also compare the performances of the different confidence and credible intervals 
of the unknown parameters in terms of their coverage percentages and credible lengths.  We further observe the performances of Algorithm 1
and Alogithm 2 by simulation experiments.

\begin{figure}[ht!]
\centering
\includegraphics[height = 6.5 cm,width=90mm]{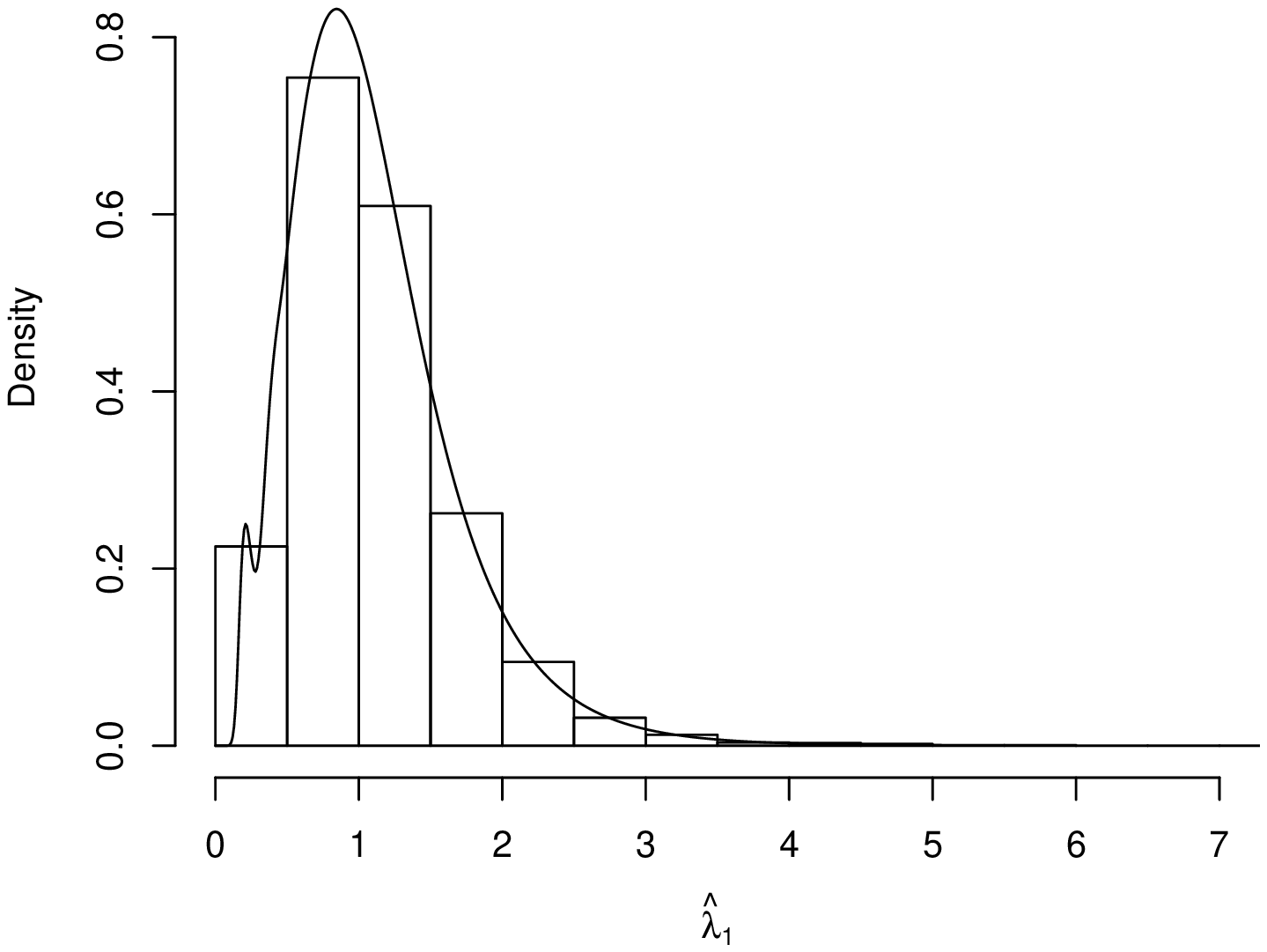}
\caption{Histogram of $\what{\lambda}_1$ along with its PDF}
\label{fig-3}
\end{figure}

\begin{figure}[ht!]
\centering
\includegraphics[height = 6.5cm, width=90mm]{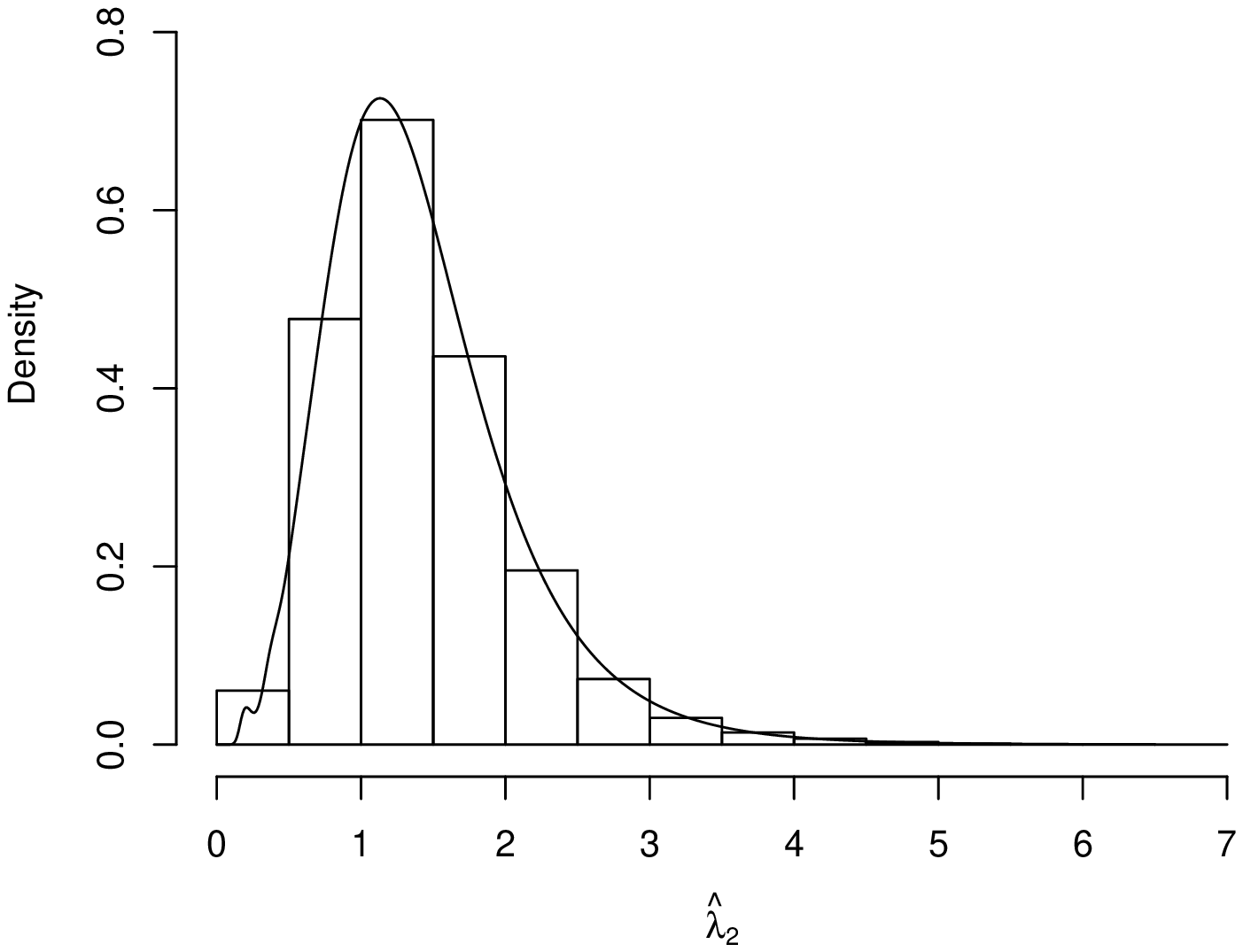}
\caption{Histogram of $\what{\lambda}_2$ along with its PDF}
\label{fig-4}
\end{figure}

In the simulation part we have taken various values of $n=10,15,20,30$ and $R=0.8n, 0.6n$ with $T=1.2$, $\lambda_1=1$ and 
$\lambda_2=1.3$.  In each case we draw random sample from the given censoring scheme and compute $\widehat{\lambda}_1$ and $\what{\lambda}_2$.  We replicate the process 
5000 times and compute the biases and MSEs.  We also compute 95\% (a) confidence intervals based on the exact distributions of 
$\widehat{\lambda}_1$ and $\what{\lambda}_2$, (b) asymptotic confidence intervals and (c) Bootstrap confidence intervals for both the parameters.  The average lengths and the coverage
percentages are computed.  All the results are reported in Table \ref{table-1}.  
In Figures \ref{fig-3} and \ref{fig-4}, we present the 
histograms and the associated PDFs of $\widehat{\lambda}_1$ and $\widehat{\lambda}_2$, given $D_1 > 0$ and $D_2 > 0$, respectively when 
$n$ = 10, $R$= 8, $\lambda_1$ = 1, $\lambda_2$ = 1.3 and $T$ = 1.2.  They match very well.

We further compute the Bayes estimates and the associated credible intervals of the unknown parameters.  
For Bayesian analysis we have taken both informative and non informative priors.  In case of informative prior, hyper parameters are 
taking 
as $\ds a_0=1.0, b_0=2.3, a_1=1.0, a_2=1.3$. These values are chosen such that the prior expectation of the parameters are exactly same as 
the corresponding true value of the parameter.  Hence, the prior is informative about the unknown parameters.  
For the non informative prior, we take $a_0=0.001, b_0=0.001, a_1=0.001, a_2=0.001$, so that 
the Bayes estimators and the proposed estimators match.  In this case the comparison between the two estimators makes sense.
For comparison purposes we compute the average biases and MSEs of the Bayes estimates under both the
priors.   We also compute 95\% credible intervals, HPD credible intervals,  and obtain the corresponding coverage percentages and average 
credible lengths over 5000 replications.  
All the results are reported in Table \ref{table-2} and Table \ref{table-3}.  

Some of the points are quite clear from these simulation experiments.  In all these cases as sample sizes increases performances of all the
estimators improve in terms of lower biases and lower MSEs.  The average lengths of the confidence intervals and credible intervals also 
decrease as sample size increases.  From Table \ref{table-1} it is clear that the performances of the confidence intervals based on 
the exact distributions of $\widehat{\lambda}_1$ and $\what{\lambda}_2$ and the bootstrap confidence intervals are quite satisfactory.   In both these cases the coverage 
percentages are very close to the nominal level.  The average lengths of the confidence intervals based on bootstrap method are smaller 
than the corresponding confidence intervals based on the exact distributions of $\widehat{\lambda}_1$ and $\what{\lambda}_2$.  The confidence intervals based on the asymptotic distribution of the MLEs are not able to maintain the associated nominal level.

From Table \ref{table-2} and Table \ref{table-3}, it is clear that  the Bayes estimates are working well.  As expected the Bayes estimates
based on the informative priors are better than the corresponding Bayes estimates based on the non-informative priors both in terms of
biases and MSEs.  In both these cases the coverage percentages of the credible intervals are very close to 
the nominal level, although in case of HPD credible intervals the coverage percentages are slightly smaller than the nominal level 
particularly when the sample size is very small.  Again, as expected the average lengths of the credible intervals based on the informative priors are smaller than 
the corresponding average lengths based on the non-informative priors.  Comparing Table \ref{table-1} and Table \ref{table-3}, it clear 
that the Bayes estimates with non-informative prior are working better than the proposed estimators.  Therefore, for all practical purposes, we recommend
to use the Bayes estimates with non-informative priors.

\begin{table}[h]
\caption{Under frequentist set up for $\lambda_1=1.0,\lambda_2=1.3, T=1.2$} \label{table-1}
\begin{center}
\begin{tabular}{|c c c c c c c|}
\hline
Sample size & Parameters & Bias & MSE & Approximate CI & Bootstrap CI & Asymptotic CI \\ [0.5ex]
\hline
n=10 & $\lambda_1$ & 0.092 & 0.341 & 2.188 (95.86) & 2.441 (93.30) & 2.080 (91.86) \\ 
 R=0.6n & $\lambda_2$ & 0.142 & 0.499 & 2.485 (95.16) & 2.848 (92.90) & 2.404 (92.17)  \\
 n=10 & $\lambda_1$ & 0.093 & 0.342 & 2.189 (95.88) & 2.440 (93.60) & 2.083 (91.91)\\ 
 R=0.8n & $\lambda_2$ & 0.151 & 0.474 & 2.495 (95.72) & 2.858 (94.08) & 2.413 (93.84)\\ [1ex]
 \hline
 n=15 & $\lambda_1$ & 0.062 & 0.211 & 1.709 (95.24) & 1.837 (93.56) & 1.654 (93.37) \\ 
 R=0.6n & $\lambda_2$ & 0.070 & 0.270 & 1.927 (95.24) & 2.106 (94.00) & 1.887 (93.38)  \\ 
 n=15 & $\lambda_1$ & 0.058 & 0.205 & 1.708 (95.42) & 1.833 (93.98) & 1.652 (92.79) \\
 R=0.8n & $\lambda_2$ & 0.086 & 0.267 & 1.941 (95.48) & 2.118 (94.18) & 1.901 (94.24) \\[1ex]
 \hline
 n=20 & $\lambda_1$ & 0.051 & 0.149 & 1.459 (95.22) & 1.539 (93.86) & 1.424 (93.14) \\ 
 R=0.6n & $\lambda_2$ & 0.061 & 0.195 & 1.652 (94.72) & 1.766 (93.54) & 1.626 (94.00)  \\ 
 n=20 & $\lambda_1$ & 0.051 & 0.148 & 1.458 (95.14) & 1.537 (93.90) & 1.424 (93.58) \\
 R=0.8n & $\lambda_2$ & 0.061 & 0.189 & 1.651 (95.54) & 1.761 (94.38) & 1.626 (94.54)\\[1ex] 
 \hline
 n=30 & $\lambda_1$ & 0.029 & 0.093 & 1.164 (94.94) & 1.206 (94.08) & 1.145 (93.80) \\ 
 R=0.6n & $\lambda_2$ & 0.044 & 0.121 & 1.326 (95.16) & 1.386 (94.52) & 1.313 (94.64)  \\ 
 n=30 & $\lambda_1$ & 0.029 & 0.092 & 1.163 (95.00) & 1.204 (94.28) & 1.144 (93.70) \\
 R=0.8n & $\lambda_2$ & 0.038 & 0.121 & 1.322 (94.66) & 1.379 (93.92) & 1.308 (93.68)  \\ [1ex]
 \hline
\end{tabular}
\end{center}
\end{table}

\begin{table}[h]
\caption{Under informative Bayesian set up for $\lambda_1=1.0,\lambda_2=1.3, T=1.2$} \label{table-2}
\begin{center}
\begin{tabular}{|c c c c c c|}
\hline
Sample size & Parameters & Bias & MSE & Credible interval & HPD Credible interval \\ [0.5ex]
\hline
n=10 & $\lambda_1$ & 0.054 & 0.188 & 1.778 (96.40) & 1.689 (94.60)  \\ 
 R=0.6n & $\lambda_2$ & 0.024 & 0.248 & 2.059 (96.54) & 1.980 (95.12)  \\
 n=10 & $\lambda_1$  & 0.054 & 0.189 & 1.778 (96.42) & 1.689 (94.12) \\ 
 R=0.8n & $\lambda_2$ & 0.032 & 0.233 & 2.065 (97.00) & 1.987 (95.48)\\ [1ex]
 \hline
 n=15 & $\lambda_1$ & 0.044 & 0.144 & 1.495 (95.76) & 1.441 (94.58)  \\ 
 R=0.6n & $\lambda_2$  & 0.046 & 0.176 & 1.709 (96.08) & 1.661 (95.16)  \\ 
 n=15 & $\lambda_1$ & 0.041 & 0.140 & 1.494 (95.98) & 1.440 (94.46) \\
 R=0.8n & $\lambda_2$ & 0.018 & 0.172 & 1.721 (96.54) & 1.673 (95.60) \\[1ex]
 \hline
 n=20 & $\lambda_1$ & 0.040 & 0.113 & 1.320 (95.44) & 1.283 (94.38) \\ 
 R=0.6n & $\lambda_2$ & 0.014 & 0.142 & 1.511 (95.36) & 1.477 (94.78)  \\ 
 n=20 & $\lambda_1$ & 0.041 & 0.112 & 1.321 (95.70) & 1.283 (94.56)  \\
 R=0.8n & $\lambda_2$  & 0.014 & 0.138 & 1.511 (96.30) & 1.477 (95.42)\\[1ex] 
 \hline
 n=30 & $\lambda_1$  & 0.024 & 0.078 & 1.090 (95.22) & 1.068 (94.46) \\ 
 R=0.6n & $\lambda_2$  & 0.015 & 0.098 & 1.250 (95.50) & 1.230 (95.08) \\ 
 n=30 & $\lambda_1$  & 0.025 & 0.077 & 1.090 (95.38) & 1.067 (94.30) \\
 R=0.8n & $\lambda_2$ & 0.010 & 0.099 & 1.247 (95.22) & 1.226 (94.54) \\ [1ex]
 \hline
\end{tabular}
\end{center}
\end{table}

\begin{table}[h]
\caption{Under non informative Bayesian set up for $\lambda_1=1.0,\lambda_2=1.3, T=1.2$} \label{table-3}
\begin{center}
\begin{tabular}{|c c c c c c|}
\hline
Sample size & Parameters & Bias & MSE & Credible interval & HPD Credible interval \\ [0.5ex]
\hline
n=10 & $\lambda_1$ & 0.091 & 0.342 & 2.025 (93.62) & 1.899 (90.94)  \\ 
 R=0.6n & $\lambda_2$ & 0.141 & 0.499 & 2.363 (93.62) & 2.251 (92.02)  \\
 n=10 & $\lambda_1$  & 0.091 & 0.344 & 2.024 (93.16) & 1.899 (90.86) \\ 
 R=0.8n & $\lambda_2$ & 0.151 & 0.474 & 2.372 (94.14) & 2.261 (93.14)\\ [1ex]
 \hline
 n=15 & $\lambda_1$ & 0.058 & 0.205 & 1.627 (93.74) & 1.558 (92.14)  \\ 
 R=0.6n & $\lambda_2$  & 0.070 & 0.270 & 1.865 (94.16) & 1.804 (93.02)  \\ 
 n=15 & $\lambda_1$ & 0.062 & 0.211 & 1.627 (93.90) & 1.559 (92.46) \\
 R=0.8n & $\lambda_2$ & 0.086 & 0.266 & 1.878 (94.90) & 1.818 (93.80) \\[1ex]
 \hline
 n=20 & $\lambda_1$ & 0.051 & 0.149 & 1.408 (93.70) & 1.363 (92.42) \\ 
 R=0.6n & $\lambda_2$ & 0.061 & 0.195 & 1.612 (93.74) & 1.572 (93.66) \\ 
 n=20 & $\lambda_1$ & 0.052 & 0.148 & 1.408 (94.12) & 1.363 (93.06)  \\
 R=0.8n & $\lambda_2$  & 0.061 & 0.189 & 1.612 (94.84) & 1.572 (94.10)\\[1ex] 
 \hline
 n=30 & $\lambda_1$  & 0.029 & 0.093 & 1.136 (94.22) & 1.112 (93.40) \\ 
 R=0.6n & $\lambda_2$  & 0.044 & 0.121 & 1.305 (94.60) & 1.282 (94.00) \\ 
 n=30 & $\lambda_1$  & 0.029 & 0.093 & 1.136 (94.36) & 1.111 (93.56) \\
 R=0.8n & $\lambda_2$ & 0.038 & 0.121 & 1.301 (94.16) & 1.278 (93.38) \\ [1ex]
 \hline
\end{tabular}
\end{center}
\end{table}

\begin{table}[h]
\caption{Under Informative prior for $ g(\lambda_1,\lambda_2) \text{with}~ \lambda_1=1.0,\lambda_2=1.3, T=1.2$} \label{table-4}
\begin{center}
\begin{tabular}{|c c c c c|}
\hline
\small{Sample size}  &   \small{Bias}		&	\small{MSE}		&	\small{Credible Interval}	&	
\small{HPD Credible Interval}  \\ [0.5ex]
\hline
n=10 \par R=0.6n &   0.002  &   0.017		&	0.514 (96.14)	&	0.507 (93.54) \\ 
 n=10 \par R=0.8n &  0.002  &   0.017 	&	0.513 (95.82)	&	0.506 (92.86)\\[1ex]
 \hline
 n=15 \par R=0.6n &  0.001  &   0.013 	&	0.447 (95.12)	&	0.443 (93.14)\\ 
 n=15 \par R=0.8n &  0.001  &   0.013 	&	0.446 (95.36)	&	0.442 (93.30)\\
 \hline
 n=20 \par R=0.6n &   0.001 &   0.010 	&	0.400 (95.42)	&	0.398 (93.88)\\ 
 n=20 \par R=0.8n &	 0.001 &	   0.010	    &   0.400 (95.56)	&	0.398 (93.86)\\ 
 \hline
 n=30 \par R=0.6n &  0.001  &   0.007		&	0.339 (95.14)	&	0.337 (93.86)\\ 
 n=30 \par R=0.8n &  0.001  &   0.007 	&	0.339 (95.50)	&	0.337 (94.48)\\
 \hline
\end{tabular}
\end{center}
\end{table}

\begin{table}[h]
\caption{Under Non Informative prior for $ g(\lambda_1,\lambda_2) \text{with}~ \lambda_1=1.0,\lambda_2=1.3, T=1.2$} \label{table-5}
\begin{center}
\begin{tabular}{|c c c c c|}
\hline
\small{Sample size}  &   \small{Bias}		&	\small{MSE}		&	\small{Credible Interval}	
&	\small{HPD Credible Interval}  \\ [0.5ex]
\hline
n=10 \par R=0.6n &   0.001  &   0.026		&	0.550 (93.54)	&	0.538 (88.34) \\ 
 n=10 \par R=0.8n &  0.001  &   0.025 	&	0.551 (93.62)	&	0.539 (88.96)\\[1ex]
 \hline
 n=15 \par R=0.6n &  0.001  &   0.018 	&	0.472 (93.96)	&	0.467 (90.14)\\ 
 n=15 \par R=0.8n &  0.001  &   0.017 	&	0.472 (93.98)	&	0.467 (90.06)\\
 \hline
 n=20 \par R=0.6n &   0.001 &   0.013 	&	0.420 (94.62)	&	0.416 (92.46)\\ 
 n=20 \par R=0.8n &	 0.001 &	   0.013    &   0.419 (93.68)	&	0.416 (91.80)\\ 
 \hline
 n=30 \par R=0.6n &  0.001  &   0.009		&	0.350 (94.06)	&	0.348 (92.78)\\ 
 n=30 \par R=0.8n &  0.001  &   0.009 	&	0.350 (94.02)	&	0.348 (92.68)\\
 \hline
\end{tabular}
\end{center}
\end{table}

\begin{table}[h]
\caption{Area and coverage percentage of Credible set for $\lambda_1=1.0,\lambda_2=1.3, T=1.2$} \label{table-6}
\begin{center}
\begin{tabular}{|c c c|}
\hline
\small{Sample size}  &   \small{Area (Informative prior)}	&	\small{Area (Non Informative prior)}  \\ [0.5ex]
\hline
n=10 \par R=0.6n &   3.877 (92.70) &   5.285 (89.40) \\ 
 n=10 \par R=0.8n &  3.890 (93.30) &   5.119 (90.26)\\[1ex]
 \hline
 n=15 \par R=0.6n &  2.755 (92.84)  &   3.207 (89.14)\\ 
 n=15 \par R=0.8n &  2.724 (92.64) &   3.253 (88.46)\\
 \hline
 n=20 \par R=0.6n &   2.073 (92.70) &   2.383 (88.56)\\ 
 n=20 \par R=0.8n & 2.085 (92.82)  &   2.384 (90.10)\\ 
 \hline
 n=30 \par R=0.6n &  1.421 (91.50) &   1.563 (89.14)\\ 
 n=30 \par R=0.8n &  1.434 (92.48) &   1.556 (89.48)\\
 \hline
\end{tabular}
\end{center}
\end{table}

Now to see how Algorithm 1 and Algorithm 2 behave we have performed some small simulation experiments and the results are 
reported in Tables \ref{table-4} to \ref{table-6}.  In all the cases we have taken $\lambda_1$ = 1.0, $\lambda_2$ = 3.0 and
$T$ = 1.2.  For Algorithm 1, we have taken $\ds g(\lambda_1, \lambda_2) = \frac{\lambda_1}{\lambda_1 + \lambda_2}$, and considered
the same informative and non-informative as discussed before.  We have computed the Bayes estimates of $\ds g(\lambda_1, \lambda_2)$
and the associated credible intervals.  The average biases, the MSEs, the average lengths of the credible intervals and the associated
coverage percentages are reported in Table \ref{table-4} and Table \ref{table-5}.  It is clear from the table values that the 
performances of Algorithm 1 are quite satisfactory.  It is observed that for informative priors HPD credible intervals perform better
than the symmetric credible intervals, where as for non-informative priors it is the other way.  Now to see the performance of 
Algorithm 2, we have computed the area of the credible set $C_{\alpha}$ and the associated coverage percentages.  The results are 
reported in Table \ref{table-6}.  It is observed as expected that the performances of the Algorithm 2 based on informative priors 
are better than the non-informative priors.

\section{\sc Data analysis}

In this section we provide the analysis of a data set for illustrative purposes.  The data are obtained from an experiment 
conducted by Dr. H.E. Walburg, Jr., of the Oak Ridge National Laboratory, see Hoel \cite{Hoel:1972}.   
This is an autopsy data for a group of RFM germ free male mice received a radiation dose of 300r at age 5-6 weeks. The original data 
are classified to be coming out from 3 causes viz. (1) Thymic Lymphoma, (2) Reticulum Cell Sarcoma, (3) Other causes. 
We have combined the two causes Thymic Lymphoma and Reticulum Cell Sarcoma as Cause-1 and other causes to be 
Cause-2.  We have made a transformation of the data $Z = (X/100)^{2.5}$, and considered a sample with $n$ = 20, $R$ = 16 and
$T$ = 5.6.  The transformed data are as follows:  0.10353(2),  0.11682(2), 0.18889(2), 0.30630(2), 3.15113(1), 3.35099(2), 4.22495(2), 4.83342(1), 4.96100(1), 5.42323(1), 5.55983(1), 5.98183(2), 6.05396(1), 7.03899(1), 7.19843(2), 7.68960(2).  Here the first figure indicates
the transformed lifetime of the mice and the associated cause of death is reported in the bracket.

\begin{figure}[ht!]
\centering
\includegraphics[height = 6.5cm, width=90mm]{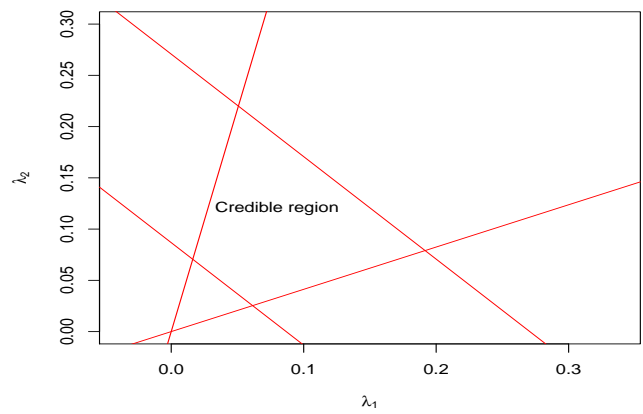}
\caption{Credible set of $\lambda_1$ and $\lambda_2$}
\label{fig-5}
\end{figure}

In this case we have, $D_1=7, D_2=9, W=\sum_{i=1}^R Z_{i:n} +(n-R)Z_{R:n}=96.94137$. The estimates of $\lambda_1$ and $\lambda_2$ are 
$\what{\lambda}_1=0.07221$ and $\what{\lambda}_2=0.09284$, respectively.  Note that they are the Bayes estimates of the corresponding parameters under non informative priors also.   Different confidence and credible intervals at 95\% level of significance are reported in Table \ref{real_data}.  In Figure \ref{fig-5} we provide the 95\% credible set of $\lambda_1$ and $\lambda_2$.

Now the natural question is whether exponential distribution provides a good fit or not to the 
above data set.  For that purpose, we have fitted the exponential distribution to the above lifetime data without the causes of 
death.  We obtain Kolmogorov-Smirnov (K-S) distance between the empirical distribution function and the estimated distribution 
function. The K-S distance is 0.28107 with the associated \textit{p} value 0.1306.  Hence, exponential distribution can be used
in this case.

\begin{table}
\caption{Confidence and credible intervals of the unknown parameters of real data}\label{real_data}

\scalebox{1.0}{

\centering
\begin{tabular}{|c| c c c c|}
\hline
&	Approximate CI	&	Bootstrap CI	&	Asymptotic CI	&	Bayes CI    \\ [0.5ex]
\hline
$\lambda_1$	&	\big(0.03027, 0.14048 \big)	&	\big(0.02957, 0.14945\big)	&	\big(0.01870, 0.12569 \big)	&	\big(0.02888, 0.13433\big) \\[1ex]
$\lambda_2$	&	\big(0.04344, 0.16699\big)	&	\big(0.04588, 0.17943\big)	&	\big(0.03218, 0.15349\big)	&	\big(0.04273, 0.16342\big)\\[1ex]	
\hline
\end{tabular}}
\end{table}

\section{\sc Conclusion}
\noindent In this paper we have considered classical and Bayesian inference of Type-II hybrid censored competing risks data.
The MLEs of the unknown parameters do not always exist.  We propose alternative estimators of the scale parameters which always exist.  They  
coincide with the MLEs, when the later exist and they are asymptotically equivalent.  We obtain the exact distribution of the proposed
estimators and based on the exact distributions confidence intervals are obtained.  For comparison purposes, we have also 
considered asymptotic and bootstrap confidence intervals.  Extensive simulation results indicate that the confidence intervals based
on the exact distribution of the proposed estimators and the bootstrap confidence intervals work quite well.  To perform the Bayesian analysis, 
we have taken very flexible Beta-Gamma distribution as a prior distribution of the unknown parameters.  The Bayes estimates and the 
associated highest posterior distributions credible intervals have been computed.  The performance of the Bayes estimates are 
quite satisfactory, and they can be implemented in practice quite easily.  Although, for notational simplicity only two causes of 
failures have been considered, the results can be easily extended for more than two causes.

In this paper for analytical simplicity we have 
assumed that the competing causes of failures follow exponential distribution.  It is well known that exponential distribution
has some serious limitations.  It will be more interesting to consider the 
case when the latent failures distribution may not be exponential.  More work is needed along that direction.

\section*{\sc Acknowledgements:} The authors would like to thank the referees, the associate editor and the editor for their 
constructive suggestions which have helped us to improve the paper significantly. 

\section*{\sc Appendix}
In the Appendix we obtain $P(\what{\lambda}_1 \le x)$
\beanno
P(\what{\lambda}_1 \leq x) & = & P(\what{\lambda}_1 \leq x| D_1=0) P(D_1=0)+ P(0 < \what{\lambda}_1 \leq x| D_1>0)P(D_1>0) \nonumber \\
& = & P(D_1=0)+ P(0 < \what{\lambda}_1 \leq x| D_1>0)P(D_1>0).
\eeanno
We derive $P(D_1=0)$ and $P(0 < \what{\lambda}_1 \leq x| D_1>0)P(D_1>0)$ separately.
\begin{align*}
P(D_1=0)&=P(D_1=0,T<Z_{R:n})+P(D_1=0,Z_{R:n}<T)\\
& = \sum\limits_{i=0}^{R-1} \binom{n}{i}
\big(1-e^{-T\lambda}\big)^i e^{-(n-i)T\lambda}
\bigg(\frac{\lambda_2}{\lambda_1+\lambda_2}\bigg)^R \\
&\quad+ \sum\limits_{i=R}^{n}
\binom{n}{i} \big(1-e^{-T\lambda}\big)^i e^{-(n-i)T\lambda}
\bigg(\frac{\lambda_2}{\lambda_1+\lambda_2}\bigg)^i .
\end{align*} 
\beanno
P(0 < \what{\lambda}_1 \leq x| D_1>0) & = & P\left[\frac{1}{\what{\lambda}_1}\geq x \big |   D_1>0\right]  \\
& = &  P \left[ \frac{1}{\what{\lambda}_1}\geq x, T\leq Z_{R:n} \big |D_1>0 \right ] + P \left[\frac{1}{\what{\lambda}_1}\geq x,Z_{R:n}<T \big |D_1>0\right] \\
& = & \sum \limits_{i=1}^R P\left[\frac{1}{\what{\lambda}_1}\geq x,T\leq Z_{R:n},D_1=i \big |D_1>0\right] \\
&  &  + \sum\limits_{j=R}^n\sum\limits_{i=1}^j P\left[\frac{1}{\what{\lambda}_1}\geq x,Z_{R:n}<T,J=j,D_1=i \big |D_1>0\right]  \\
& = & \sum\limits_{i=1}^R P\left[\frac{1}{\what{\lambda}_1}\geq x \big |T<Z_{R:n},D_1=i\right]P \left [T<Z_{R:n},D_1=i \big |D_1>0 \right ]\\
&  & +\sum\limits_{j=R}^n\sum\limits_{i=1}^j P\left[\frac{1}{\what{\lambda}_1}\geq x \big |Z_{R:n}<T,J=j,D_1=i\right]\\
& & \quad P \left [Z_{R:n}<T,J=j,D_1=i \big |D_1>0 \right ] \\
& = & \sum\limits_{i=1}^R G_i(x)q_i+\sum\limits_{j=R}^n\sum\limits_{i=1}^jG_{ij}(x)q_{ij},
\eeanno
where
\noindent
\begin{align*}
G_i(x)&= P\Big[\frac{1}{\what{\lambda_1}}\geq x \big |T<Z_{R:n},D_1=i\Big], &q_i=P\Big[T<Z_{R:n},D_1=i \big |D_1>0\Big], \\
G_{ij}(x)&=P\Big[\frac{1}{\what{\lambda}_1}\geq x \big |Z_{R:n}<T,J=j,D_1=i\Big], &q_{ij}=P\Big[Z_{R:n}<T,J=j,D_1=i \big |D_1>0\Big].\\
\end{align*}
We provide each of the above terms separately.  For $i=1,2,\ldots,n$
\noindent
\begin{align}
\noindent q_i &=P\left[T<Z_{R:n},D_1=i \big |D_1>0\right] \nonumber\\
&= \frac{1}{P[D_1>0]}P \left [T<Z_{R:n}\big]P\big[D_1=i|T<Z_{R:n} \right ]\nonumber \\
&=\displaystyle \frac{1}{P[D_1>0]}P\big(T<Z_{R:n}\big){R\choose i} \bigg(\frac{\lambda_1}{\lambda_1+\lambda_2}\bigg)^i\bigg(\frac{\lambda_2}{\lambda_1+\lambda_2}\bigg)^{R-i}. \label{eq1}\nonumber\\
\end{align}
The last term of equation (\ref{eq1}) holds since, given $T<Z_{R:n}$, $D_1 \sim$ Bin\bigg(R,$\frac{\lambda_1}{\lambda_1+\lambda_2}$\bigg).

For $j=R,\ldots,n$, we note that $J \sim$ Bin$ \big(n,1-e^{-T\lambda}\big)$ \hspace{1mm}\text{and} $D_1 \sim $ Bin$\bigg(j,\frac{\lambda_1}{\lambda_1+\lambda_2}\bigg)$. Thus for $j=R,\ldots,n$ \hspace{0.2mm} and \hspace{0.2mm}  $i=1,\ldots, j$ we have, 
\noindent
\begin{align}
q_{ij}&=P\Big[Z_{R:n}<T,J=j,D_1=i|D_1>0\Big]\nonumber\\
&=\frac{1}{P[D_1>0]}P\Big[Z_{R:n}<T,J=j,D_1=i\Big]\nonumber\\
&=\frac{1}{P[D_1>0]}P\Big[J=j,D_1=i\Big] \quad \text{\{since, for $j=R,\ldots,n, \{J=j\} \implies  \{Z_{R:n}<T\}$\} }\nonumber\\
&=\frac{1}{P[D_1>0]} P\Big[J=j\big]P\Big[D_1=i|J=j\Big]\nonumber\\
&=\frac{1}{P[D_1>0]} {n\choose j} \Big(1-e^{-T\lambda}\Big)^j e^{-T(n-j)\lambda} {j \choose i}\bigg(\frac{\lambda_1}{\lambda_1+\lambda_2}\bigg)^i \bigg(\frac{\lambda_2}{\lambda_1+\lambda_2}\bigg)^{j-i}. \nonumber\\
\end{align}
To find $G_i(x)$ and $G_{ij}(x)$ we need the following two lemmas.

\noindent {\sc Lemma 2:} The joint distribution of $Z_{1:n},\ldots,Z_{R:n}$ given $T<Z_{R:n}$, $D_1=i$,
for $i=1,\ldots,R$ at $z_1 < \ldots < z_R$, is given by
$$
f_{Z_{1:n},\ldots, Z_{R:n}|T<Z_{R:n},D_1=i}(z_1,\ldots, z_R) = \frac{1}{P(T<Z_{R:n})} \frac{n!}{(n-R)!}  \lambda^R
e^{-\lambda \sum\limits_{k=1}^R z_k} e^{-\lambda (n-R)z_R}.
$$
\noindent {\sc Proof:}  Let us consider 
\beanno
&  & P(z_1<Z_{1:n}<z_1+dz_1,\ldots, z_R<Z_{R:n}<z_R+dz_R|T<Z_{R:n},D_1=i)\\
&  & = \frac{1}{P(T<Z_{R:n},D_1=i)} P(z_1<Z_{1:n}<z_1+dz_1,\ldots,z_R<Z_{R:n}<z_R+dz_R,T<Z_{R:n},D_1=i)\\
&  & =\frac{\lambda_1^i \lambda_2^{R-i} 
e^{-\lambda \sum\limits_{k=1}^R z_k} e^{-\lambda (n-R)z_R}\dbinom{R}{i}\frac{n!}{(n-R)!}}{P(T<Z_{R:n})P(D_1=i|T<Z_{R:n})}dz_1\ldots dz_R \\
&  & =\frac{1}{P(T<Z_{R:n})} \frac{n!}{(n-R)!} \lambda^R e^{-\lambda \sum\limits_{k=1}^R z_k} e^{-\lambda (n-R)z_R} dz_1\ldots dz_R.  
\eeanno 
\qed

\noindent {\sc Lemma 4:} The joint distribution of $Z_{1:n},\ldots,Z_{J:n}$ given $Z_{R:n}<T,D_1=i,J=j$
for $i=1,2,\ldots,j$, $j=R,\ldots,n$, at $z_1 < \ldots < z_J$, is given by
$$
f_{Z_{1:n},\ldots, Z_{J:n}|Z_{R:n}<T,D_1=i,J=j}(z_1,\ldots, z_J)=\frac{j!\lambda^je^{-\lambda\sum\limits_{k=1}^jz_k}}{(1-e^{-\lambda})^j}.
$$
\noindent {\sc Proof:} Let us consider
\beanno
&  & P \left (z_1<Z_{1:n}<z_1+dz_1,\ldots, z_J<Z_{J:n}<z_J+dz_J \big |Z_{R:n}<T,D_1=i,J=j \right )\\
&  & =\frac{P(z_1<Z_{1:n}<z_1+dz_1,\ldots, z_J<Z_{J:n}<z_J+dz_J,Z_{R:n}<T,D_1=i,J=j)}{P(Z_{R:n}<T,D_1=i,J=j)}\\
&  & = \frac{\lambda_1^i \lambda_2^{j-i} 
e^{-\lambda\sum\limits_{k=1}^jz_k} 
e^{-(n-j)T\lambda} \frac{n!}{(n-j)!} \dbinom{j}{i}}{\dbinom{n}{j}(1-e^{-T\lambda})^je^{-T\lambda (n-j)}\dbinom{j}{i}(\frac{\lambda_1}{\lambda_1+\lambda_2})^i (\frac{\lambda_2}{\lambda_1+\lambda_2})^{j-i}} dz_1,\ldots,dz_j\\
&  & =\frac{j! \lambda ^je^{-\lambda \sum\limits_{k=1}^jz_k}}{(1-e^{-T\lambda})^j} dz_1,\ldots,dz_j.    
\eeanno
\qed

To find $G_i(x)$, we find conditional MGF $E(e^{t\frac{1}{\what{\lambda}_1}}|T<Z_{R:n},D_1=i)$ for $i=1,\ldots,R$ and is derived below.
\bea
&  & E\left[e^{t\frac{1}{\what{\lambda}_1}}|T<Z_{R:n},D_1=i\right] \nonumber \\
& = & E\left[e^{t\frac{\sum\limits_{k=1}^RZ_{k:n}+(n-R)Z_{R:n}}{D_1}} \biggr| T<Z_{R:n},D_1=i\right]  \nonumber \\
& = & E\left[e^{\frac{t}{i} (\sum\limits_{k=1}^{R-1} Z_{k:n}+(n-R+1)Z_{R:n})} \biggr|T<Z_{R:n},D_1=i\right]  \nonumber \\
& = & \frac{1}{P(T<Z_{R:n})} \frac{n!}{(n-R)!} \lambda^R   \nonumber \\
& & \ \ \ \
\int\limits_T^\infty\ \int\limits_0^{z_R}\ldots \int\limits_0^{z_2}
e^{-\big(\lambda-\frac{t}{i}\big)z_1-\ldots \big(\lambda-\frac{t}{i}\big)z_{R-1}}
e^{-\big(n-R+1\big)\big(\lambda-\frac{t}{i}\big)z_R} \hspace{0.2mm}dz_1 \ldots\hspace{0.2mm}dz_{R-1}\hspace{0.2mm} dz_R  \nonumber \\
\eea
\bea
& = & \frac{1}{P(T<Z_{R:n})} \frac{n!}{(n-R)!} \lambda^R  \left (\lambda-\frac{t}{i} \right )^{1-R}\frac{1}{(R-1)!} \nonumber \\
& & \ \ \ \
\int\limits_T^\infty \bigg[1-e^{-(\lambda-\frac{t}{\i}\big)z_R}\bigg]^{R-1} \bigg [ \int\limits_0^{z_R}\ldots \int\limits_0^{z_2} (R-1)! \frac{(\lambda-\frac{t}{i})^{R-1}}{\bigg[1-e^{-(\lambda-\frac{t}{\i}\big)z_R}\bigg]^{R-1}} \times \bigg .  \nonumber \\
&  & \ \ \ \quad \quad \quad \quad \quad \bigg .
e^{-\big(\lambda-\frac{t}{i}\big)z_1-\ldots \big(\lambda-\frac{t}{i}\big)z_{R-1}}dz_1 \ldots\hspace{0.2mm}dz_{R-1} \bigg ] 
 e^{-\big(n-R+1\big)\big(\lambda-\frac{t}{i}\big)z_R} \hspace{0.2mm}\hspace{0.2mm} dz_R  \nonumber \\
& & \text{(Note that the above $R-1$ fold integration is 1 as it is the PDF of}   \\  \nonumber 
&  & \ \ \ \ \text{ truncated order statistics)}\nonumber\\
& = & \frac{1}{P(T<Z_{R:n})} \frac{n!}{(n-R)!} \lambda^R \bigg(\lambda-\frac{t}{i}\bigg)^{1-R}
\frac{1}{(R-1)!}  \times \nonumber  \\
&  & \int\limits_T^\infty \bigg[1-e^{-\big(\lambda-\frac{t}{\i}\big)z_R}\bigg]^{R-1} e^{-\big(n-R+1\big)\big(\lambda-\frac{t}{i}\big)z_R}\hspace{0.2mm}dz_R  \nonumber  \\
& = & \frac{1}{P(T<Z_{R:n})} \frac{n!}{(n-R)!(R-1)!} \lambda^R \bigg(\lambda-\frac{t}{i}\bigg)^{1-R}
     \times \nonumber \\
&  & \sum\limits_{k=0}^{R-1} \binom{R-1}{k} (-1)^k \int\limits_T^\infty e^{-\big(\lambda-\frac{t}{i}\big)\big(n-R+1+k\big)z}
\hspace{0.1mm}dz  \nonumber  \\
& =  & \frac{1}{P(T<Z_{R:n})} n \binom{n-1}{R-1} \bigg(1-\frac{t}{i\lambda}\bigg)^{-R}\hspace{2 mm} \sum\limits_{k=0}^{R-1} \binom{R-1}{k} (-1)^k \hspace{2 mm}\frac{e^{-\big(\lambda-\frac{t}{i}\big)\big(n-R+1+k\big)T}}{\big(n-R+1+k\big)}.   \label{cond-mgf}
\eea
Hence, using the uniqueness property of the MGF, from (\ref{cond-mgf}), we can obtain the conditional PDF of 
$\ds \frac{1}{\widehat{\lambda}_1}$
given that $T < Z_{R:n}$ and $D_1 = i$ for $i=1,2,\ldots R$, as the following,
\begin{align*}
f_{\frac{1}{\what{\lambda}_1}|D_1=i,T<Z_{R:n}}\big(x\big)&=\frac{1}{P(T<Z_{R:n})} n \binom{n-1}{R-1} \sum\limits_{k=0}^{R-1} \frac{\binom{R-1}{k}(-1)^k}{n-R+1+k} e^{-T\lambda \left(n-R+1+k\right)}\\
& \hspace{50mm}f_{G}\left(x;\frac{T}{i}\left(n-R+1+k\right),R,i\lambda\right).
\end{align*}
Therefore, $G_i(x)$ can be obtained as
$$
G_i(x) = \int_x^\infty f_{\frac{1}{\what{\lambda}_1}|D_1=i,T<Z_{R:n}}\big(y\big)dy.
$$
Similarly to find $G_{ij}(x)$ for $i=1,\ldots,j$ and $j=R,\ldots,n$, we find conditional moment generating function
of $E(e^{t\frac{1}{\hat{\lambda}_1}}|Z_{R:n}<T,D_1=i,J=j)$  and is derived below.
\begin{align*}
&E\bigg[e^{\frac{t}{\widehat{\lambda}}_1} \biggr |Z_{R:n}<T,D_1=i,J=j\bigg]\\
&=E\bigg[e^{\frac{t}{i}\sum\limits_{k=1}^jZ_{k:n}+(n-j)T} \biggr |Z_{R:n}<T,D_1=i,J=j\bigg]\\
&=\frac{j!\lambda^j}{(1-e^{-\lambda})^j} e^{\big(n-j\big)\frac{Tt}{i}}\int\limits_0^T\ \int\limits_0^{z_j}\ldots \int\limits_0^{z_2} e^{-\big(\lambda-\frac{t}{i}\big)z_1-\ldots-\big(\lambda-\frac{t}{i}\big)z_{j-1}-\big(\lambda-\frac{t}{i}\big)z_j} \hspace{0.2mm}dz_1\ldots dz_{j-1}dz_j\\
&=\frac{j!\lambda^j}{(1-e^{-T\lambda})^j(j-1)!} e^{\big(n-j\big)\frac{Tt}{i}} \bigg(\lambda-\frac{t}{i}\bigg)^{1-j} \int\limits_0^T e^{-\big(\lambda-\frac{t}{i}\big)z} \bigg[1-e^{-\big(\lambda-\frac{t}{i}\big)z}\bigg]^{j-1}\hspace{2 mm} dz\\
&=\frac{e^{(n-j)\frac{Tt}{i}}}{(1-e^{-T\lambda})^j} \bigg(1-\frac{t}{i\lambda}\bigg)^{-j} \bigg[1-e^{-\big(\lambda-\frac{t}{i}\big) T}\bigg]^j  \times \\
& \ \ \ \ \ \ \ \frac{1}{(1-e^{-T\lambda})^j} \sum\limits_{k=0}^j \binom{j}{k} (-1)^k e^{-\big(\lambda-\frac{t}{i}\big)Tk} e^{(n-j)\frac{Tt}{i}} \bigg(1-\frac{t}{i\lambda} \bigg)^{-j}.
\end{align*}
Hence, similarly as before,  we can obtain the conditional PDF of $\frac{1}{\widehat{\lambda}_1}$
given that $T < Z_{R:n}$, $D_1 = i$ and $J = j$, for $i=1,2,\ldots, j$ and $j=R,R+1,\ldots,n$ as the following.
\begin{align*}
f_{\frac{1}{\widehat{\lambda}_1} \big |Z_{R:n}<T,J=j,D_1=i}\big(x\big)&=\frac{1}{(1-e^{-T\lambda})^j} \sum\limits_{k=0}^j \binom{j}{k} (-1)^k e^{-Tk\lambda} f_{G}\bigg(x;\big(n-j+k\big)\frac{T}{i},j,i\lambda\bigg).
\end{align*}
Therefore, $G_{ij}(x)$ can be obtained as
$$
G_{ij}(x) = \int_x^\infty f_{\frac{1}{\widehat{\lambda}_1} \big |Z_{R:n}<T,J=j,D_1=i}\big(y\big)dy.
$$

\noindent {\sc Proof of Lemma 5:}

Note that to prove Lemma 5, it is enough to prove that for any $b > 0$, 
$$
\lim_{\lambda_1 \rightarrow 0} P_{\lambda_1}(\widehat{\lambda}_1 \le b) \rightarrow 1 \ \ \ \hbox{and} \ \ \  
\lim_{\lambda_1 \rightarrow \infty} P_{\lambda_1}(\widehat{\lambda}_1 \le b) \rightarrow 0.
$$
For $b > 0$, the distribution function of $\what{\lambda}_1$ is written as before
\beanno
P_{\lambda_1}(\what{\lambda}_1 \leq b) & = & \sum \limits_{i=0}^{n} c_i(\lambda_1) + \sum \limits_{i=1}^R \sum \limits_{s=0}^{R-1} 
c_{i,s}(b;\lambda_1,\lambda_2) + \sum \limits_{j=R}^n \sum \limits_{i=1}^{j} \sum \limits_{s=0}^{j} d_{j,i,s}(b;\lambda_1,\lambda_2), 
\eeanno
where for fixed $b$ and $\lambda_2$,\\
\noindent (a) $c_i(\lambda_1)= c_i$ as a function of $\lambda_1$ for $i=0,1,\ldots, n$

\noindent (b) $c_{i,s}(b;\lambda_1,\lambda_2)= c_{i,s}(b)$ as a function of $\lambda_1$ for $i=1,2,\ldots,R$; $s=0,1,\ldots, R-1$

\noindent (c) $d_{j,i,s}(b;\lambda_1,\lambda_2)= d_{j,i,s}(b)$ as a function of $\lambda_1$ for $s=0,\ldots,j$; $i=1,\ldots,j$ and $j=R,\ldots,n$.
We consider two cases separately.

\noindent {\sc Case-I} $\lambda_1 \rightarrow 0$

\noindent Note that for $i=0,1,\ldots, n$,
\beanno 
c_i(\lambda_1) \rightarrow \binom{n}{i} (1-e^{-T\lambda_2})^i e^{-(n-i)T\lambda_2} \quad \text{as} \quad \lambda_1 \rightarrow 0 
\eeanno
Thus $\sum_{i=0}^n c_i(\lambda_1) \rightarrow 1$ as $\lambda_1 \rightarrow 0.$

For $i=1,\ldots, R \quad \text{and} \quad s=0,\ldots, R-1$,
\beanno
c_{i,s} (b;\lambda_1,\lambda_2) \rightarrow 0 \quad \text{as} \quad \lambda_1 \rightarrow 0.
\eeanno

For $s=0,\ldots,j,\quad  i=1, \ldots, j\quad  \text{and}\quad  j=R,\ldots, n,$
\beanno
d_{j,i,s} (b;\lambda_1,\lambda_2) \rightarrow 0 \quad \text{as} \quad \lambda_1 \rightarrow 0.
\eeanno
Hence $P_{\lambda_1}(\hat{\lambda}_1 \leq b) \rightarrow 1$ as $\lambda_1 \rightarrow 0$ for fixed $b > 0$ and $\lambda_2$.

\noindent {\sc Case-II} $\lambda_1 \rightarrow \infty$

\noindent Note that for $i=0,1,\ldots, n$,
\beanno 
c_i(\lambda_1) \rightarrow 0 \quad \text{as} \quad \lambda_1 \rightarrow \infty. 
\eeanno
For $i=1,\ldots, R \quad \text{and} \quad s=0,\ldots, R-1$,
\beanno
c_{i,s} (b;\lambda_1,\lambda_2) \rightarrow 0 \quad \text{as} \quad \lambda_1 \rightarrow \infty .
\eeanno
Note that,
\[
\lim_{\lambda_1 \rightarrow \infty} d_{j,i,s}(b;\lambda_1,\lambda_2)=\begin{dcases*}
\lim_{\lambda_1 \rightarrow \infty} \overline{F}_G \bigg(\frac{1}{b};n,n (\lambda_1+ \lambda_2) \bigg)	&	if $s=0; i=j=n$\\
0	&	o.w.
\end{dcases*}		
\]
Now,
\begin{align*}
\overline{F}_G \bigg(\frac{1}{b};n,n (\lambda_1+ \lambda_2)\bigg)& = \frac{\bigg(n(\lambda_1+\lambda_2)\bigg)^n}{\Gamma(n)} \int_{\frac{1}{b}}^\infty e^{-n(\lambda_1+\lambda_2)z} z^{n-1} dz\\[9pt]
& =\frac{1}{\Gamma(n)}\int_{\frac{n(\lambda_1+\lambda_2)}{x}}^\infty e^{-u} u^{n-1} du \quad \quad \text{taking,}~u=n(\lambda_1+\lambda_2)z\\[9pt]
&	\rightarrow 0 \quad \quad \text{as} \quad \lambda_1 \rightarrow \infty.
\end{align*}
Hence $\ds P_{\lambda_1}(\hat{\lambda}_1 \leq b) \rightarrow 0$ as $\lambda_1 \rightarrow \infty$ for fixed $b > 0$ and $\lambda_2$.

Since $\ds P_{\lambda_1}(\what{\lambda}_1 \leq b)$ is a monotonically decreasing and continuous function of $\lambda_1$, the solutions of 
two equations $\ds P_{\lambda_1}(\what{\lambda}_1 \leq b)= \frac{\alpha}{2}$ and $\ds P_{\lambda_1}(\what{\lambda}_1 \leq x) 
= 1-\frac{\alpha}{2}$, always exist.


\end{document}